\documentclass[12pt]{article}
\usepackage[left=2cm,top=2cm,right=2cm,bottom=2cm]{geometry}
\usepackage{authblk}

\usepackage{appendix}
\usepackage[cp1251]{inputenc}
\usepackage[english]{babel}
\usepackage{xcolor}
\usepackage{cite}

\usepackage{amsfonts,graphicx,epsfig,epstopdf}
\usepackage{amssymb,amsmath,enumerate,float,amsthm}
\usepackage{stmaryrd}
\usepackage{array}
\usepackage{hyperref}

\usepackage[scr=boondox,cal=euler]{mathalfa}
\usepackage{mdframed} 

\renewcommand{\arraystretch}{1.5}
\usepackage{hyperref}
\hypersetup{
    colorlinks=true,
    linkcolor=blue,
    citecolor = magenta,
    }

\makeatother
\graphicspath{{./figures/}}
\theoremstyle{plain}

\newtheorem*{analogy}{Analogy}
\newtheorem*{Greensprop}{Discrete Green's identity}

\theoremstyle{definition}

\newenvironment{rema}
{%
	\pushQED{\qed}\begin{rema/}}
	{\popQED\end{rema/}}

\newtheorem{defi/}[theorem]{Definition}
\newtheorem{rema/}[theorem]{Remark}
\newtheorem{exa/}[theorem]{Example}

\addto\captionsenglish{}

\newcommand{\ptl}{\partial}

\newcommand{\be}{\begin{equation}}
\newcommand{\ee}{\end{equation}}
\newcommand{\beq}{\begin{equation}}
\newcommand{\eeq}{\end{equation}}

\newcommand{\lx}{s}
\newcommand{\ly}{q}
\newcommand{\lk}{{\tilde k}}
\newcommand{\bdxi}{\boldsymbol{\xi}}
\newcommand{\blx}{\boldsymbol{\lx}}
\newcommand{\rmU}{{ \boldsymbol{\rm \Psi}}}

\newcommand{\rmA}{{\boldsymbol{\rm K}}}
\newcommand{\rmF}{{ \boldsymbol{\rm F}}}
\newcommand{\gF}{\boldsymbol{\mathcal{F}}\!}
\newcommand{\gK}{\boldsymbol{\mathcal{K}}\!}

\newcommand{\kstar}{\xi^{\rm in}}
\newcommand{\bn}{{\boldsymbol n}}
\newcommand{\bnu}{{\boldsymbol{\nu}}}
\newcommand{\bmu}{{\boldsymbol{\mu}}}
\usepackage{enumitem}
\setlist[enumerate]{label*=\arabic*.}

\newcommand{\RED}[1]{{\color{black}{#1}}} 
\newcommand{\teal}[1]{{\color{black}{#1}}} 
\newcommand{\green}[1]{{\color{black}{#1}}} 

\title{\green{On an analogy between the Wiener--Hopf formulations of discrete and continuous diffraction problems}}
\author{A. I. Korolkov, R. C. Assier, A. V. Kisil}
\begin{document}
\maketitle

\begin{abstract}

\teal{This article is dedicated to unifying the framework used to derive the Wiener--Hopf equations arising from some discrete and continuous wave diffraction problems.
The main tools are the discrete Green's identity and the appropriate notion of discrete normal derivative. The resulting formal analogy between the Wiener--Hopf equations allows one to effortlessly move between the discrete and continuous formulations. The validity of this novel analogy is illustrated through several famous two-dimensional canonical diffraction problems and extended to three-dimensional problems.}

\end{abstract}

\section{Introduction}

 \teal{Discrete diffraction problems on lattices have been recently successfully studied with the Wiener--Hopf method (see e.g.~\cite{Sharma2015,maurya2019scattering,Kisil2024,medvedeva2024diffraction, Nieves_2d_24}) and more generally with analytical methods~\cite{Shanin2020,Shanin2022,Bhat_Osting_09,Martin_06,Craster_16}. Such discrete problems can be interpreted as numerical approximations of their continuous counterparts~\cite{Finite_difference_book}, but can also be considered as spring-mass systems \cite{Mass_spring_review}. They are used, for example, in fracture mechanics~\cite{slepyan2012models} and as a basis for more complicated lattice constructions~\cite{Nieves_21,Mishuris_beam, MovchanBook}. Hence, the topic and results of the present work are also relevant to these areas, but we will concentrate on diffraction problems here.

 \RED{Despite being initially designed to solve a specific type of integral equation \cite{WienerHopf1931}, the Wiener--Hopf method~\cite{WHreviewAK} has proven to be a powerful technique to solve diffraction problems involving obstacles with sharp edges or corners~\cite{bookWH,LawrieAbrahams2007}. It turns the boundary value problem at hand into a \green{functional} equation with known analytical properties called a Wiener--Hopf equation.}
 
 The aim of the present article is to establish an analogy between discrete and continuous problems, allowing one to effortlessly transfer known \teal{functional equations in} classical diffraction theory to the discrete setting.

There are at least two common approaches to derive Wiener--Hopf equations for wave \RED{diffraction} problems. \RED{Jones'} method~\cite{bookWH} \RED{consists} \green{of} \RED{applying the Fourier transform to the governing equation (typically the Helmholtz equation) and} \green{the} \RED{boundary conditions on the surface of the scatterer}. A more recent approach is based on the application of Green's identity to the scattered field and an outgoing \RED{plane} wave \RED{depending} on a spectral parameter \cite{Shanin2003}.   Both approaches are essentially equivalent, however, we find the second one to be more suitable for the task of establishing an analogy with problems on lattices. 


The main difficulty in constructing a Wiener--Hopf equation for a discrete problem, using a method developed for the corresponding continuous problem, lies in replacing the notion of \RED{normal} derivative. In the present work, this problem is solved by introducing an appropriate definition, which is then used to derive a discrete version of Green's identity. We then employ this identity to derive the discrete Wiener-Hopf equations and establish the aforementioned analogy. However, one should note that the discrete normal derivative introduced \green{here} can also be used to implement a discrete version of Jones' method, which would result in the same Wiener-Hopf equations.

Moreover, we believe that the discrete version of Green's identity is also of more general interest as it can be used in other analytical methods.
For example, the continuous version of Green's identity is frequently used to derive the global relation in Fokas' unified transform method (see e.g.\ \cite{Naqvi_Ayton_25,MFokas_14}). Hence, the present work should also be helpful to formulate the global relation for discrete problems.}

\RED{The rest of the article is organised as follows. In Section \ref{sec:Green}, we remind the reader about Green's identity in the continuous case, and construct an equivalent identity for the discrete case. Section \ref{sec:dispersion} is dedicated to the analysis of the dispersion relations associated with the continuous and discrete Helmholtz equations. In Section \ref{sec: half_plane}, we briefly recall the continuous and discrete solutions to the two-dimensional half-plane problem (also known as the Sommerfeld problem) through the Wiener-Hopf method. Based on the observed similarities between the discrete and continuous form of the Wiener-Hopf equations for the half-plane problem, in Section \ref{sec:analogy}, we propose the analogy that represents the main result of this work. In Section \ref{sec:examples}, we consider several famous two-dimensional canonical diffraction problems in their continuous and discrete forms, we reduce both forms to some Wiener--Hopf equations and show that our analogy is indeed satisfied. Section \ref{sec:3D} is dedicated to an extension of this work to three-dimensional diffraction problems and in particular to the quarter-plane problem. Concluding remarks are provided in Section \ref{sec:conclusion}.}

\section{Green's identity} \label{sec:Green}

\RED{In this section}, we recall the well-known Green's identity for the Helmholtz equation and propose its discrete analogue.  

\subsection{Continuous problems}

\begin{figure}[htbp!]
    \centering
    \includegraphics[width=0.8\linewidth]{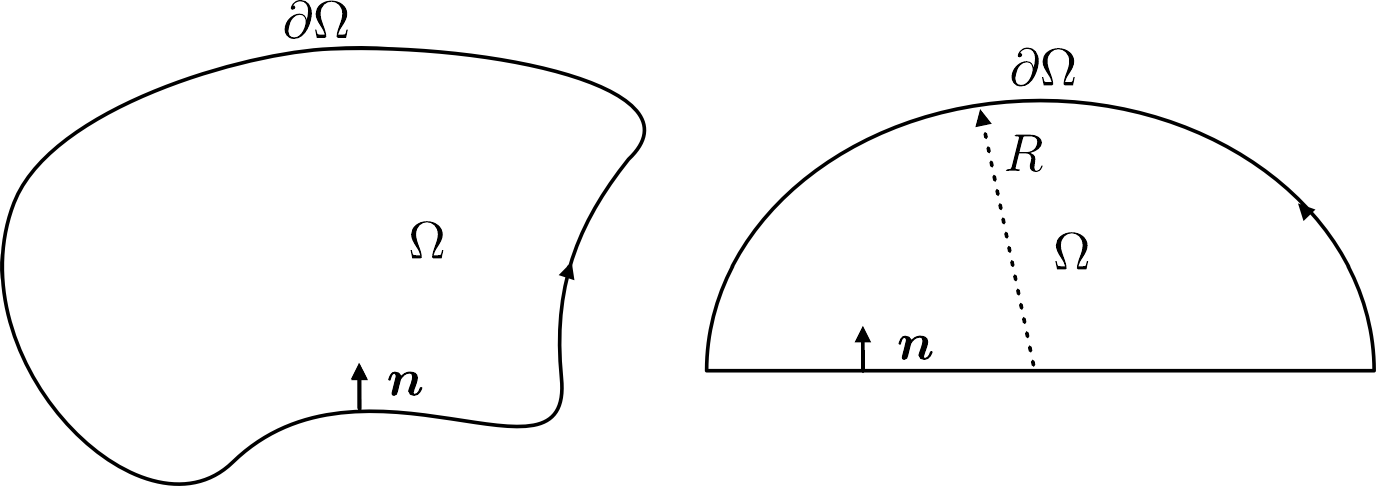}
   \caption{\green{A typical domain for Green's identity (left) and a semi-disc domain that tends to the upper half-space as $R\to\infty$} }
    \label{fig:Green_dom_cont}
\end{figure}

Let us consider two functions $u$ and $w$ that satisfy the 2D Helmholtz equation in a \green{compact} domain $\Omega$:
\[
\Delta u(x,y) + k^2 u(x,y) = f(x,y),\quad \Delta w(x,y) + k^2 w(x,y) = g(x,y),  
\]
where $k$ is a wavenumber parameter and $f$ and $g$ are given forcing terms.  Then the following  (second Green's) identity is true:
\begin{equation}
\label{eq:Green_cont_first}
\int_{\ptl \Omega}\left[u\frac{\ptl w}{\ptl \bn}-w\frac{\ptl u}{\ptl \bn}\right] dl= \int_\Omega(fw - gu)dS,
\end{equation}
where  $\ptl \Omega$ is the boundary of the domain which is passed in the positive direction, see Figure~\ref{fig:Green_dom_cont}, left, $\bn$ is the unit inward normal on $\ptl \Omega$, \RED{and the notation $\partial/\partial\bn~\equiv~\bn\cdot \nabla$ is used for the normal derivative operator.}
\begin{rema}
\label{rem:continuousgreennoncompact}
Green's identity can be generalised \RED{to} \green{unbounded} domains $\Omega$. \RED{This has to be done through a limiting procedure, considering bigger and bigger closed domains. Moreover, it is necessary to assume that the integral over the boundary parts that tend to infinity vanishes. For example, in Section~\ref{sec: half_plane}, Green's identity is applied to the upper half-space with $u$ taken as the scattered field that satisfies the radiation conditions, and $w$ being an outgoing wave with an exponential decay in the upper half-space. To do this, we first consider a semi-disc $\Omega$ of radius $R$  as shown in Figure~\ref{fig:Green_dom_cont}, right, and then we take the limit as $R\to \infty$. Because of the  exponential decay, the integral on the arc tends to zero, and only an integral along the $x$-axis (the boundary of the upper-half space) remains.}       
\end{rema}

\subsection{Lattice problems}
\begin{figure}[htbp!]
    \centering
    \includegraphics[width=0.8\linewidth]{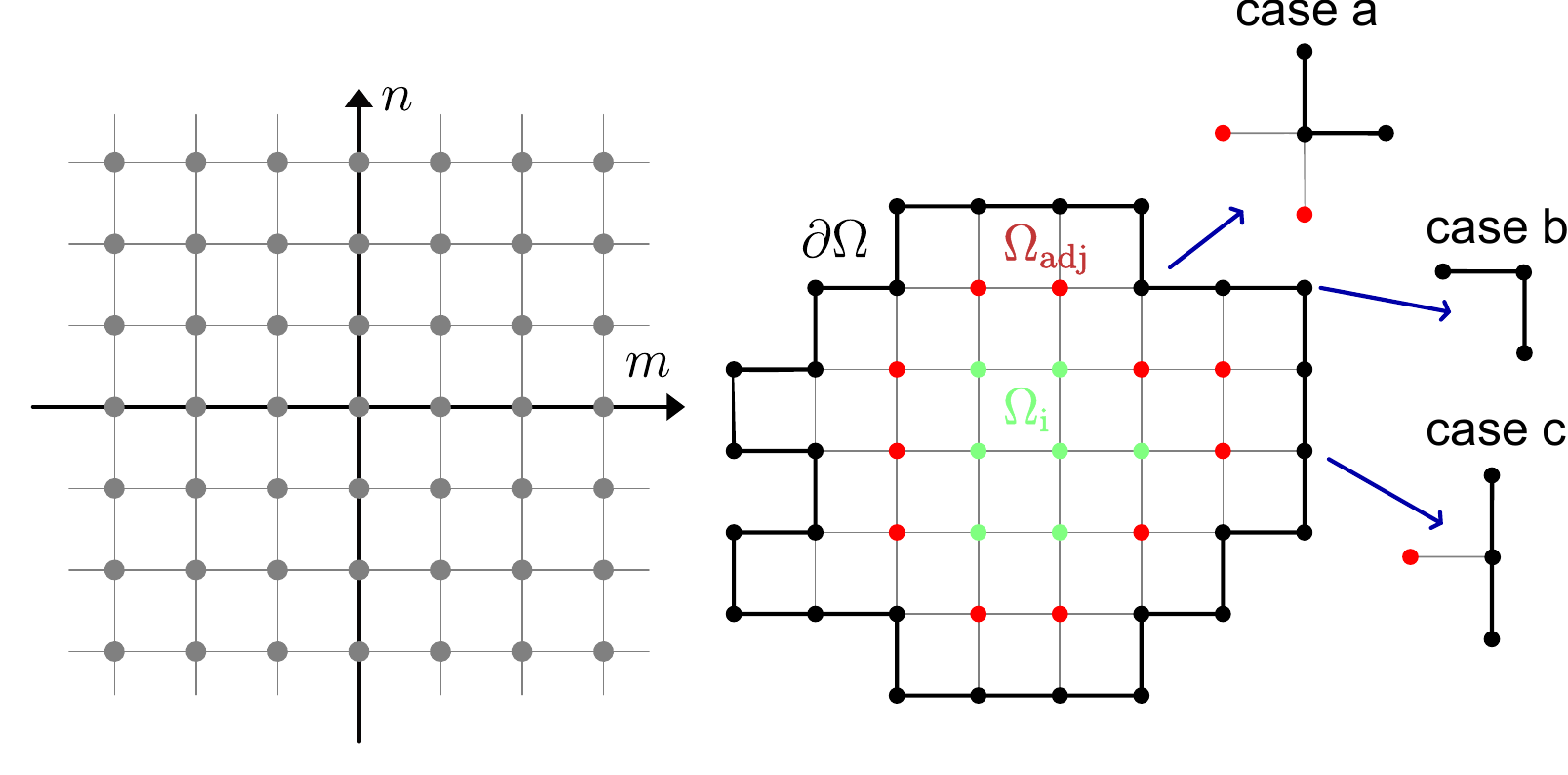}
    \caption{A typical square lattice (left) and a typical domain for Green's identity on lattices (right)}
    \label{fig:latticeGreensdomain}
\end{figure}

Let us consider a uniform square lattice numbered with \teal{integer} numbers $(m,n)$, see Figure~\ref{fig:latticeGreensdomain}, left. Consider a compact domain $\Omega$ with boundary $\ptl \Omega$, as shown in Figure~\ref{fig:latticeGreensdomain}, right. Suppose that $\Omega$ does not include $\ptl \Omega$, i.e. $\Omega\cap\ptl\Omega=\RED{\emptyset}$. 
Two nodes connected by an edge are said to be \emph{adjacent}. The set of nodes that are adjacent to a boundary node is denoted $\Omega_{\text{adj}}$.
Denote the set of all nodes in $\Omega$ that are not adjacent to the boundary by $\Omega_{\rm i}$. Thus, $\Omega = \Omega_{\rm adj}\cup\Omega_{\rm i}$. 
The introduced notations are illustrated in Figure~\ref{fig:latticeGreensdomain},~right.

Consider a pair of lattice functions $u(m,n)$ and $w(m,n)$ that satisfy a discrete analogue of the Helmholtz equation in $\Omega$:
\green{
\begin{equation}
\Delta_{(m,n)}[u] + \lk^2u(m,n) = f(m,n),\quad
\Delta_{(m,n)}[w] + \lk^2w(m,n) = g(m,n),
\label{eq:discreteg}
\end{equation}
where $\lk=kh$, $h$ being the length of the lattice edges and $k$ being the wavenumber. The operator $\Delta_{(m,n)}[\cdot]$ is the 2D discrete Laplace operator, which is a 5-point finite difference approximation of the continuos Laplace operator:
\[
\Delta_{(m,n)}[u] = u(m,n+1) + u(m,n-1) + u(m+1,n) + u(m-1,n)-4u(m,n).
\]
} 
To shorten notations, let us introduce a single multi-index ${\bnu} = (m,n)$ that numbers the nodes of the lattice\RED{, and use the notations $u_\bnu\equiv u(m,n)$, $f_{\bnu}\equiv f(m,n)$, etc...} Below, we are going to use both notations interchangeably. Then, (\ref{eq:discreteg}) can be rewritten as follows: 
\begin{equation}
\label{eq:disc_Helm}
\sum_{{\bmu}\in\Omega\cup\ptl\Omega}\beta_{{\bnu}{\bmu}}u_{{\bmu}} = f_{{\bnu}},
\quad \sum_{{\bmu}\in\Omega\cup\ptl\Omega}\beta_{{\bnu}{\bmu}}w_{{\bmu}} = g_{{\bnu}}, \quad {\bnu} \in \Omega
\end{equation}
where $\beta_{{\bnu}{\bmu}}$ is defined as follows:
\[
\beta_{{\bnu}{\bmu}}=\begin{cases}
1, & {\bnu}\neq {\bmu}, \text{ ${\bnu}$ is adjacent \RED{to} ${\bmu}$}\\
\lk^2 - 4, & {\bnu} = {\bmu},\\
0, & \text{otherwise}.
\end{cases}
\]
Let us  introduce a key notion of an analogue of the normal derivative for the boundary nodes:
\begin{equation}
\label{eq:disc_derivative}
\ptl_{\bnu}[u] = \sum_{{\bmu} \in \ptl \Omega\cup\Omega_{\rm adj}} \alpha_{{\bnu}{\bmu}} u_{\bmu} ,\quad {\bnu}\in \ptl\Omega  
\end{equation}
where $\alpha_{{\bnu}{\bmu}}$ is defined differently for different parts of the boundary. There are three possible cases (see Figure~\ref{fig:latticeGreensdomain}, right): 
\begin{itemize}
\item ${\bnu}$ lies on an external right angle, see Figure~\ref{fig:latticeGreensdomain}, case a. Then:
\[
\alpha_{{\bnu}{\bmu}}=\begin{cases}
1/2, & {\bnu}\neq {\bmu}, \text{ ${\bnu}$ is adjacent \RED{to} ${\bmu}$ and ${\bmu} \in \ptl \Omega$  }\\
1, & {\bnu}\neq {\bmu}, \text{ ${\bnu}$ is adjacent \RED{to} ${\bmu}$ and ${\bmu} \in \Omega_{\rm adj}$}\\
3(\lk^2/4 - 1), & {\bnu} = {\bmu},\\
0, & \text{otherwise}.
\end{cases}
\]
\item ${\bnu}$ lies on an internal right angle, see Figure~\ref{fig:latticeGreensdomain}, case b. Then:
\begin{equation}
\label{eq:lat_bound_Neum_b}
\alpha_{{\bnu}{\bmu}}=\begin{cases}
1/2, & {\bnu}\neq {\bmu}, \text{ ${\bnu}$ is adjacent \RED{to} ${\bmu}$}\\
\lk^2/4- 1, & {\bnu} = {\bmu},\\
0, & \text{otherwise}.
\end{cases}
\end{equation}
\item ${\bnu}$ lies on a straight line, see Figure~\ref{fig:latticeGreensdomain}, case c. Then:
\[
\alpha_{{\bnu}{\bmu}}=\begin{cases}
1/2, & {\bnu}\neq {\bmu}, \text{ ${\bnu}$ is adjacent \RED{to} ${\bmu}$ and ${\bmu} \in \ptl \Omega$  }\\
1, & {\bnu}\neq {\bmu}, \text{ ${\bnu}$ is adjacent \RED{to} ${\bmu}$ and $ {\bmu} \in \Omega_{\rm adj}$}\\
2(\lk^2/4 - 1), & {\bnu} = {\bmu},\\
0, & \text{otherwise}.
\end{cases}
\]
\end{itemize}
\green{
We are now ready to formulate a discrete analogue of (\ref{eq:Green_cont_first}).

\begin{mdframed}
\begin{Greensprop}
Let a pair of lattice functions $u_{\bnu}$ and $w_\bnu$  satisfy the discrete Helmholtz equations (\ref{eq:disc_Helm}) in a compact domain $\Omega$. Then 
\begin{equation}
\label{Green's_lattice}
\sum_{{\bnu} \in \ptl \Omega} \left(\ptl_{\bnu}[u] w_{\bnu}  - \ptl_{\bnu}[w] u_{\bnu}\right) =    \sum_{{\bnu} \in \Omega}\left(g_{\bnu} u_{\bnu} - f_{\bnu} w_{\bnu}\right).
\end{equation}
\end{Greensprop}
\end{mdframed}}
We will refer to the latter as Green's identity on a square lattice. Note, that slightly different equalities were derived in \cite{PobletPuig2018}. Identity (\ref{Green's_lattice}) will significantly simplify the derivation of functional equations for the problems we study below, and provide a straightforward analogy with the continuous case.

Let us prove (\ref{Green's_lattice}). First, by definition we have the symmetry
\[
\beta_{\bnu\bmu} = \beta_{\bmu\bnu}
\]
and the equality 
\[
\beta_{\bnu\bmu} = \alpha_{\bnu\bmu}, \quad \text{$\bnu \in \ptl \Omega$ and $\bmu\in \Omega_{\rm adj}$.}
\]

\begin{figure}[htbp!]
    \centering
    \includegraphics[width=0.4\linewidth]{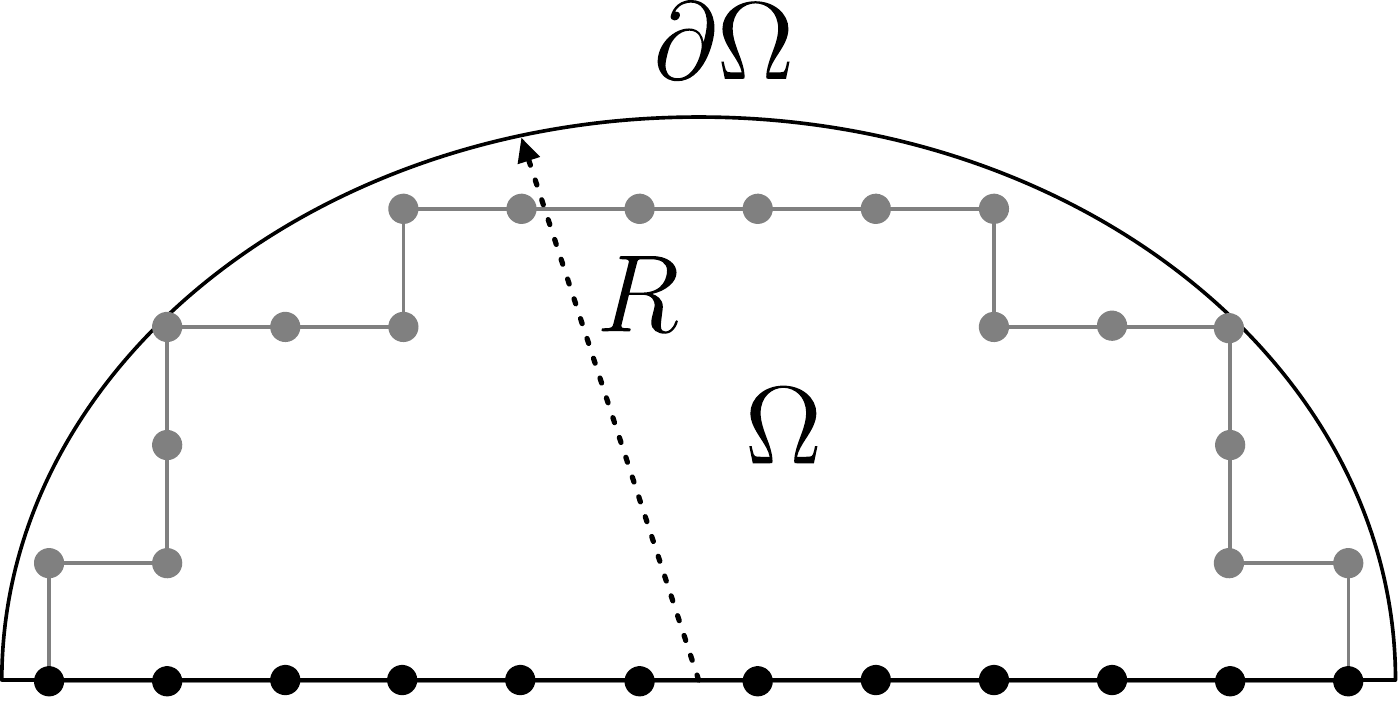}
    \caption{Domain for Green's theorem in the upper half-plane}
    \label{fig:latticeGreenhalfplane}
\end{figure}

Then, let us introduce $\bar \beta_{\bnu\bmu}$ that is equal to $\beta_{\bnu\bmu}$ for $\bnu \in \Omega$, and to $\alpha_{\bnu\bmu}$ for $\bnu\in \ptl\Omega$. Then, from the symmetry 
$
\bar \beta_{\bnu\bmu} =  \bar \beta_{\bmu\bnu}
$
follows the  identity:
\[
\sum_{\bnu \in \Omega\cup\ptl\Omega}\sum_{\bmu \in \Omega\cup\ptl\Omega}\left(w_{\bnu}\bar\beta_{\bnu\bmu}u_{\bmu}  -u_{\bnu}\bar\beta_{\bnu\bmu} w_{\bmu}\right)=0,
\]
or equivalently
\[
\sum_{\bnu \in \ptl\Omega}\sum_{\bmu \in \Omega_{\text{adj}}\cup\ptl\Omega}\left(w_{\bnu}\alpha_{\bnu\bmu}u_{\bmu}  -u_{\bnu}\alpha_{\bnu\bmu} w_{\bmu}\right) =\sum_{\bnu \in \Omega}\sum_{\bmu \in \Omega\cup\ptl\Omega}\left(u_{\bnu}\beta_{\bnu\bmu} w_{\bmu} - w_{\bnu}\beta_{\bnu\bmu}u_{\bmu}\right).
\]
Taking into account (\ref{eq:disc_Helm}) and (\ref{eq:disc_derivative}), we obtain (\ref{Green's_lattice}).
\begin{rema}
Similarly to the continuous case \RED{(see Remark \ref{rem:continuousgreennoncompact})}, one can apply Green's identity to unbounded domains, provided that the functions $u$ and $w$ decay on the part of the boundary that tends to infinity. Indeed, let us consider the important example of $\Omega$ being the upper half-space. Let $u,w$ tend to zero exponentially as $(m,n)$ tend to infinity in the upper half-space.  In Section~\ref{sec: half_plane}, the decay is provided by the limiting absorption principle, which is implemented by adding a small imaginary part to the wavenumber $\lk$. \RED{The limiting procedure is done by considering} a contour shown in Figure~\ref{fig:latticeGreenhalfplane}. The upper ``arc'' of the contour is parametrised by the parameter $R$. Studying the limit as $R\to \infty$, we obtain an identity in the upper half-space with summation only along the line $m=0$. 

Note that for some specific values of $\lk$, lattice problems can become ``degenerate'', and radiation conditions cannot be formulated at all \cite{Kapanadze2021,Shanin2024}. We exclude such \RED{degeneracies} from our consideration. 
\end{rema}

\begin{rema}
The choice of $\alpha_{{\bnu}{\bmu}}$ is not unique. However, the one we made can be considered as a numerical approximation for the normal derivative as $h\to 0$. It can be shown using the finite element method technique. Indeed, using linear shape functions to approximate the field, one will obtain the bulk equations (\ref{eq:disc_Helm}) inside the domain, and equations
\begin{equation}
\label{eq: disc_boundary_eq}
\ptl_{\bnu} [u]  = 0
\end{equation}
for the boundary nodes, if a Neumann boundary value problem is considered. The details are provided in Appendix~\ref{app:FEM}.
\end{rema}
\begin{rema}
For Green's identity to hold, we only need the symmetry of the coefficients $\bar \beta_{\bnu\bmu}$. Thus, we are not restricted to the finite difference stencils introduced above, or to square meshes. For example,  Green's identity can be generalised to a wide range of finite element stencils, rectangular or triangular lattices, and also for three-dimensional lattices, see Section~\ref{sec:3D} for the latter.      
\end{rema}
\section{Dispersion relation}\label{sec:dispersion}
To apply Green's identity, we need to have two functions which satisfy a given Helmholtz equation, one of these functions will be the unknown scattered field, and the second will be the auxiliary function we have to construct. These auxiliary functions will be taken as plane waves, hence, it is important to study which plane waves are permissible, and this is done via the dispersion relation. The dispersion relation defines a surface, \RED{each point of which} corresponding to a plane wave \RED{that} can propagate in free space. For this reason, the first step towards establishing our \emph{method of analogies} will be to study the dispersion relations arising from both continuous and discrete problems. Understanding the structure of the dispersion relation is one of the key steps towards the solution of any wave problem. 
\subsection{Continuous problems}
\begin{figure}[htbp!]
    \centering
    \includegraphics[width=0.7\linewidth]{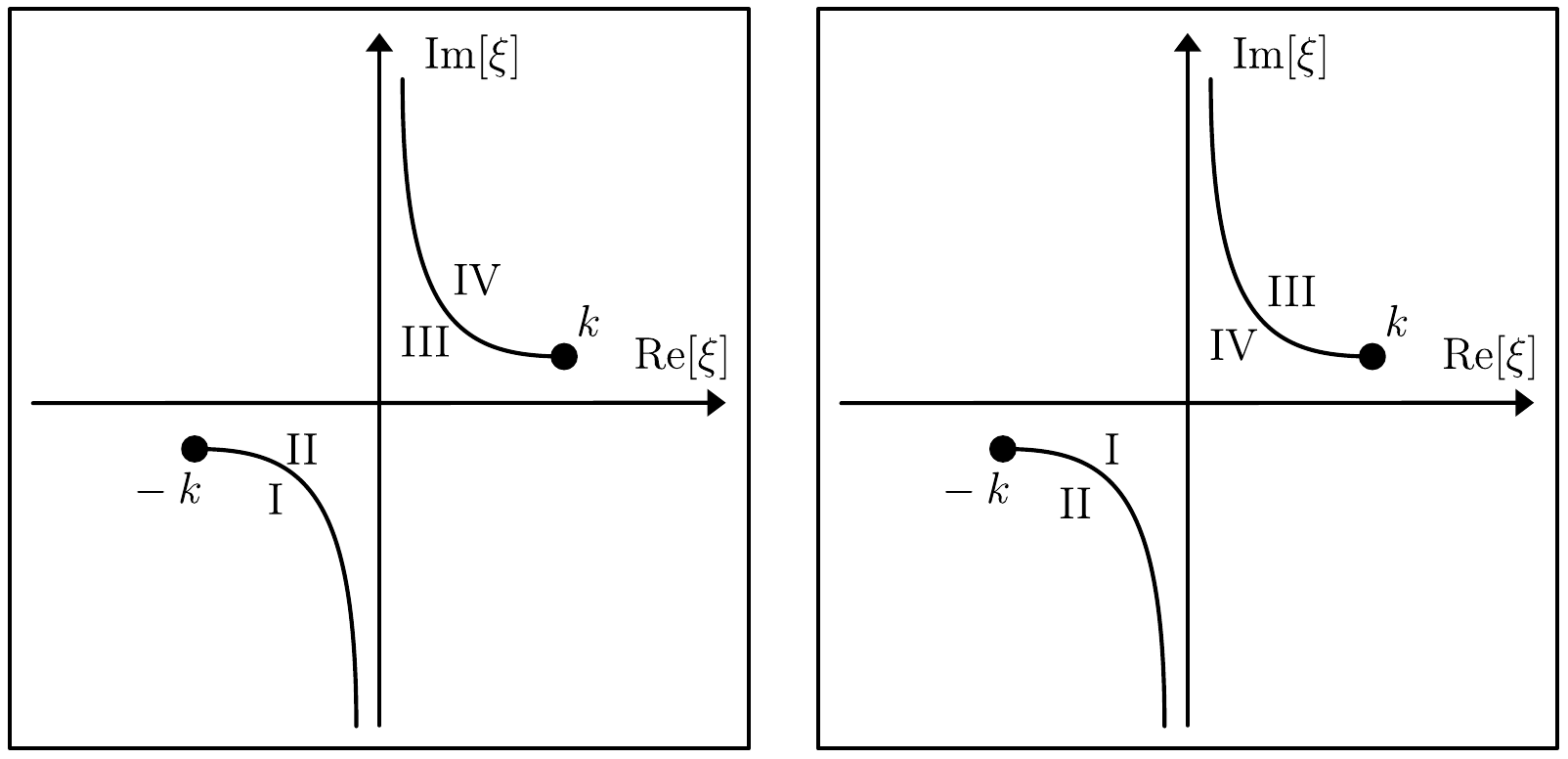}
    \caption{\RED{The} Riemann surface \RED{of} $\gamma(\xi)$. \RED{The Roman numerals describe how the cuts are glued to each other.}}
    \label{fig:gamma_Rsurf}
\end{figure}

\RED{Consider} the homogeneous Helmholtz equation 
\begin{equation}
\label{eq:Helm}
(\Delta +k^2)u(x,y) = 0,
\end{equation}
where $k$ is the wavenumber.
\RED{Let us seek plane wave solutions of the form}
\[
u(x,y) = \exp\{i\xi x + i\gamma y \},
\]
where $\xi,\gamma$ are some complex numbers.  Substituting the latter into the Helmholtz equation, we get \RED{the dispersion relation}
\[
D_c(\xi,\gamma) = -\xi^2 - \gamma^2 +k^2 = 0.
\]
The latter defines a surface in $ (\xi,\gamma) \in \mathbb{C}^2$, \RED{referred to as the} dispersion surface. 
Solving it with respect to $\gamma$ for some fixed $\xi$ we get:
\[
\gamma(\xi) = \sqrt{k^2 - \xi^2}.
\]
The function $\gamma(\xi)$ is two-valued, with a two-sheeted Riemann surface shown in Figure~\ref{fig:gamma_Rsurf}.   It has two branch points at $\xi = \pm k$. The cuts are conducted in a way that ${\rm Im}[\gamma]$ is zero on the cuts. Let us call the physical sheet of the surface the sheet on which  ${\rm Im}[\gamma]>0$.

\begin{figure}[htbp!]
    \centering
    \includegraphics[width=0.8\linewidth]{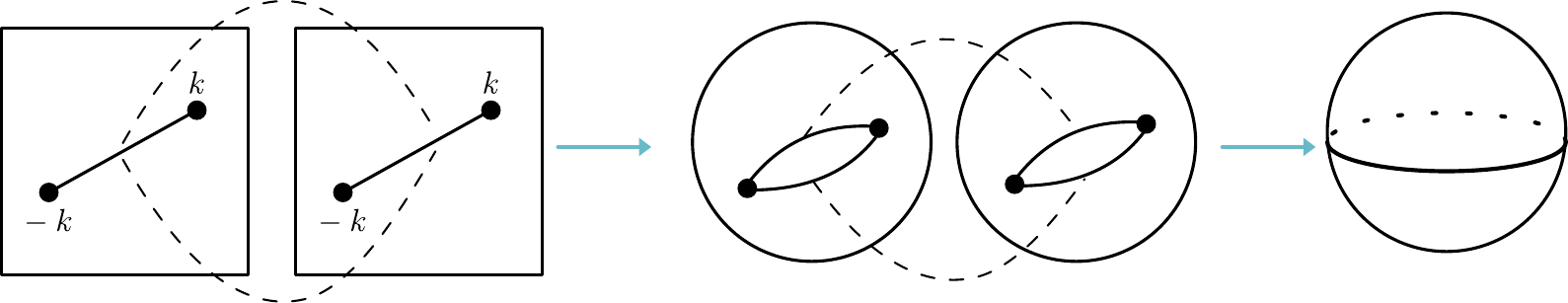}
    \caption{The process of deformation of the dispersion surface into a Riemann sphere}
    \label{fig:cont_compact}
\end{figure}

The dispersion surface has the topology of a Riemann sphere, i.e. its genus is $0$. Indeed, let us \RED{instead} choose the cuts as shown in Figure~\ref{fig:cont_compact}, and compactify each sheet into a Riemann sphere with a cut. Then, by joining the two spheres along the cuts, we obtain a single Riemann sphere.
\subsection{Lattice problems}
\begin{figure}[htbp!]
	\centering{
	\includegraphics[width=0.7\textwidth]{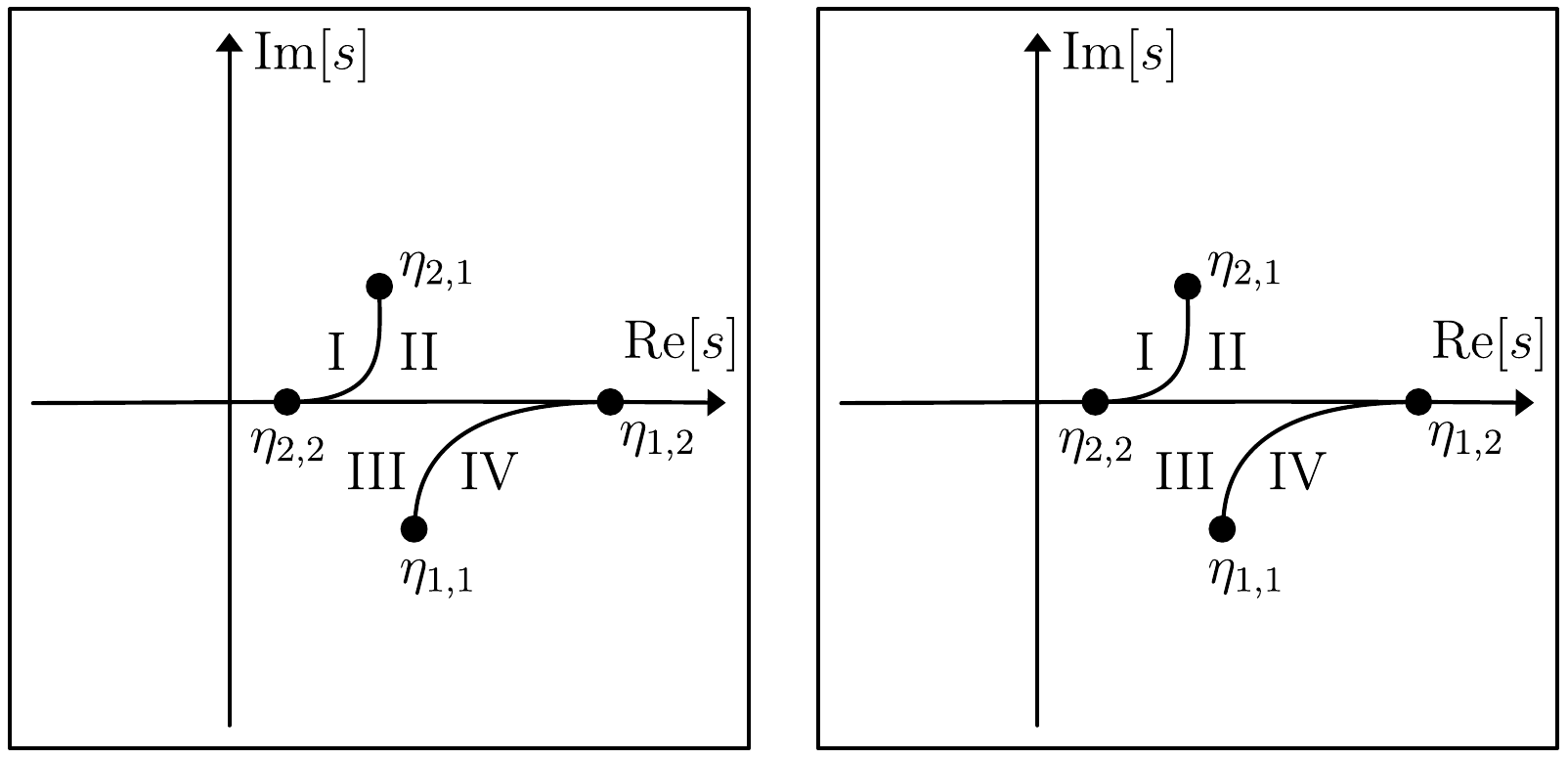}
	}
	\caption{\RED{The} Riemann surface of $\ly(\lx)$. \RED{The Roman numerals describe how the cuts are glued to eachother.}}
	\label{fig:Disp_surf_lattice}
\end{figure}

\RED{Consider} the homogeneous discrete Helmholtz equation
\begin{equation}
\label{eq:Helm_lat}
u(m,n+1) + u(m,n-1) + u(m+1,n) + u(m-1,n) + (\lk^2-4)u(m,n) = 0.
\end{equation}
\RED{Let us seek solutions of the form}
\begin{equation}
\label{eq:lattice_plane_wave}
u(m,n) = \lx^m \ly^n.
\end{equation}  
where $\lx, \ly$ are some complex numbers. Substituting the latter  into the discrete Helmholtz equation, we get 
\begin{equation}
\label{eq:lattice_Disp_eq}
D_d(\lx,\ly) \equiv \lx+ \lx^{-1} + \ly+\ly^{-1} + \lk^2-4 = 0. 
\end{equation}
We will refer to the latter as the dispersion \RED{relation} of a square lattice, and to $\lx^m \ly^n$ for $(\lx,\ly)$ chosen such that $D_d(\lx,\ly)=0$
as a plane wave on a square lattice. 
The equation (\ref{eq:lattice_Disp_eq}) can be solved with respect to $\ly$ for some fixed $\lx$:
\begin{equation}
\ly(\lx) = -\frac{\lk^2 - 4 + \lx + \lx^{-1}}{2} + \frac{\sqrt{\left(\lk^2 - 4 + \lx + \lx^{-1}\right)^2 - 4}}{2} \cdot
\end{equation}
\RED{As in the continuous case, the function $\ly(\lx)$ is two-valued and its Riemann surface is two-sheeted. However, as illustrated in Figure~\ref{fig:Disp_surf_lattice},} the function $\ly(\lx)$ now has four branch points \RED{given by}
\[
\eta_{1,1} = -\frac{d_1}{2} - \frac{i\sqrt{4-d_1^2}}{2}, \quad \eta_{2,1}  = -\frac{d_1}{2} +\frac{i\sqrt{4-d_1^2}}{2}, \quad d_1 = \lk^2 - 2, 
\]
\[
\eta_{1,2} = -\frac{d_2}{2} + \frac{\sqrt{d_2^2-4}}{2}, \quad \eta_{2,2}  = -\frac{d_2}{2} -\frac{\sqrt{d_2^2-4}}{2}, \quad d_2 = \lk^2 - 6. 
\]

Note that $\ly(\lx)$ is single valued at the branch points with  $\ly(\eta_{2,1}) = \ly(\eta_{1,1})=1$, $\ly(\eta_{2,2}) = \ly(\eta_{1,2})=-1$. Let the cuts be conducted in such a way that $|\ly|=1$ on the cuts. Let us call the physical sheet of the Riemann surface the sheet on which $|\ly|<1$ and the value of the function on that sheet by \(\ly(\lx)\). A direct calculation shows that $\ly(\lx)\ly_1(\lx)=1$, where $\ly_1(\lx)$ is the value on the non-physical sheet.

Let us introduce the function $\Upsilon(\lx)$ defined by
\[
\Upsilon(\lx) = \frac{\ly(\lx)-\ly^{-1}(\lx)}{2} = \frac{1}{2}\sqrt{\left(\lk^2-4+\lx+\lx^{-1}\right)^2-4}.
\]
It can be equivalently expressed as 
\[
\Upsilon(\lx)  = (2\lx)^{-1}\sqrt{\left(\lx - \eta_{1,1}\right)\left(\lx - \eta_{1,2}\right)\left(\lx - \eta_{2,1}\right)\left(\lx - \eta_{2,2}\right)}.
\]
As we will demonstrate below, $\Upsilon(\lx)$ acts as a lattice analogue of $\gamma(\xi)$.

Similarly to the continuous dispersion surface, one can compactify the Riemann surface of $q(s)$ into two Riemann spheres with two cuts \green{each}, and then by joining the cuts, show that \green{the resulting surface} has the topology of a torus. 
\section{Diffraction by a  half-plane}
\label{sec: half_plane}
\teal{To make the analogy apparent, we present the simplest possible  example:  diffraction by a half-plane. Based on this instructive example, the analogy will be formulated in Section \ref{sec:analogy}.}
\subsection{The continuous problem}
\begin{figure}[htbp!]
	\centering{
	\includegraphics[width=0.8\textwidth]{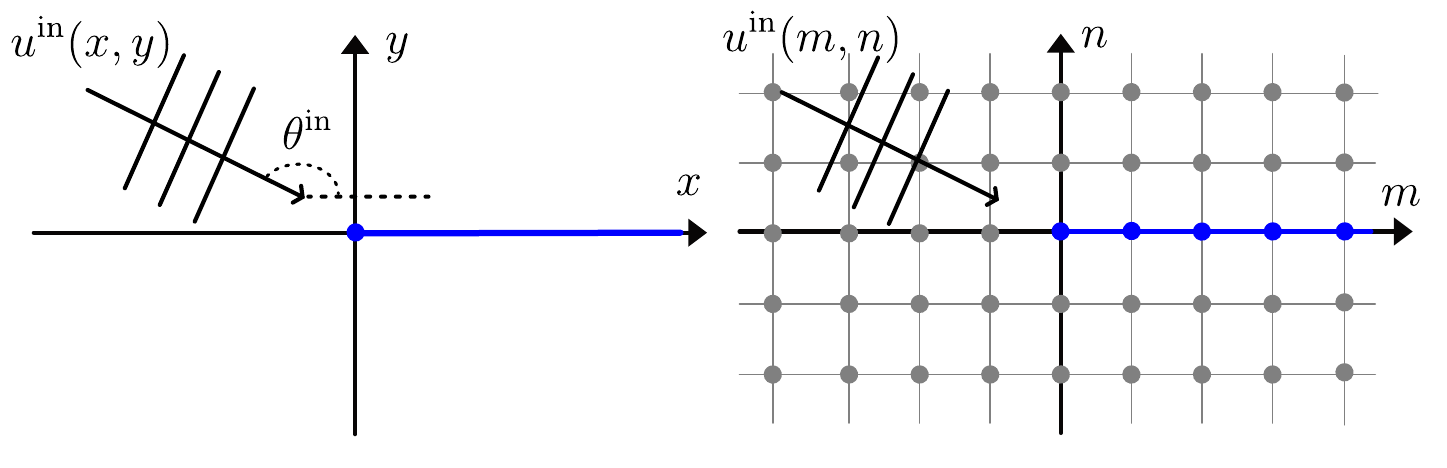}
	}
	\caption{Geometries of the continuous (left) and lattice (right) problems of diffraction by a half-plane}
	\label{fig:half_plane}
\end{figure}

Consider the classical problem of diffraction of an incident plane wave by a half-plane with Dirichlet boundary conditions. The geometry of the problem is shown in Figure~\ref{fig:half_plane}, left. 
Let the total field $u$ satisfy the homogeneous Helmholtz equation (\ref{eq:Helm})
everywhere except on the half-line where we have
\[
u(x,0) = 0, \quad x\geq 0.
\]
Let us write the total field as the sum of the incident wave $u^{\rm in}$ and \RED{the} scattered \RED{field} $u^{\rm sc}$:
\begin{equation}
\label{eq:inc_field}
u(x,y) = u^{\rm in} + u^{\rm sc}, \quad u^{\rm in}(x,y) = \exp\{-ikx\cos\theta^{\rm in} - iky\sin\theta^{\rm in}\}.
\end{equation}
The radiation condition is imposed in terms of the limiting absorption principle, i.e.\ we assume that $k$ has a small positive imaginary part\footnote{In the context of time-dependent problem it means that we are choosing \RED{the} $\exp\{-i\omega t\}$ \RED{convention}.}. 
\teal{For the problem to have a unique solution, the total field should also satisfy Meixner condition at the edge}
\begin{equation}
\label{Meixner_cond}
u^{\rm sc} = Cr^{1/2}\sin\left(\frac{\theta}{2}\right) + O(r^{3/2}) \RED{\text{ as } r\to 0},
\end{equation}
where $x = r\cos\theta$, $y=r\sin\theta$.
Using a symmetry argument \cite{bookWH}, the problem can be restricted to the  upper half-space with mixed boundary conditions on the line $x=0$:
\[
\frac{\ptl u^{\rm sc}}{\ptl y}(x,0) = 0 \text{ \RED{for} } x<0 \quad \text{\RED{and}} \quad u^{\rm sc}(x,0) = - \exp\{-ikx\cos\theta^{\rm in}\} \text{ \RED{for} } x\geq0.
\]
Motivated by the dispersion relation, let us introduce \RED{an} auxiliary  function $w$ defined by
\begin{equation}
\label{eq:w_fun}
w(x,y) = \exp\{i\xi x  + i\gamma(\xi)y\},\quad \gamma(\xi) = \sqrt{k^2 - \xi^2}.
\end{equation}
It decays exponentially in the upper half-space and can be treated as an outgoing plane wave. 
Since $w$ and $u^{\rm sc}$ both satisfy the homogeneous Helmholtz equation (\ref{eq:Helm}), we can apply the second Green's identity to get
\begin{equation}
\label{eq: Greens_identity}
\int_{\ptl \Omega }\left[w\frac{\ptl u^{\rm sc}}{\ptl \bn}-u^{\rm sc}\frac{\ptl w}{\ptl \bn}\right]dl\RED{=\int_{-\infty}^{\infty} \left[ w \frac{\ptl u^{\rm sc}}{\ptl y}-u^{\rm sc} \frac{\ptl w}{\ptl y}\right] dx}  =0,
\end{equation}
where \RED{we have applied the limiting procedure described in Remark \ref{rem:continuousgreennoncompact} and illustrated in Figure~\ref{fig:Green_dom_cont}, right. Using the boundary condition and the definition of $w$, it leads to the functional equation}

\begin{equation}
\label{eq: Wh_half_plane}
 \Psi_c^-(\xi)=K_c(\xi)\Psi_c^+ + F_c(\xi),   
\end{equation}
where
\[
K_c(\xi)  = \frac{1}{i\gamma(\xi)}, \quad F_c(\xi) = \frac{i}{\xi - \kstar},\quad \kstar = k\cos\theta^{\rm in},
\]
 and $\Psi_c^-$ and $\Psi_c^+$ are unknown spectral functions (one-sided Fourier transforms) defined by: 
\begin{equation}
\label{eq:a_coef}
\Psi_c^-(\xi)  = \int_{-\infty}^{0} \exp\{i\xi x\}u^{\rm sc}(x,0) dx,\quad \Psi_c^+(\xi)  = \int_{0}^{\infty} \exp\{i\xi x\}\frac{\ptl u^{\rm sc}}{\ptl y}(x,0) dx.
\end{equation}
The latter defines the functions $\Psi_c^-(\xi)$ and $\Psi_c^+(\xi)$ analytical in the lower and the upper half-planes, respectively. Thus (\ref{eq: Wh_half_plane}) is a Wiener-Hopf equation on the line $(-\infty,\infty)$ \RED{with kernel $K_c$ and forcing term $F_c$}. 

The behaviour of these unknown spectral functions at infinity can be inferred from Meixner's conditions via Watson's lemma~\cite{bookWH}. Indeed, it follows from (\ref{Meixner_cond}) that 
\[
\Psi_c^{-}(\xi) = O(\xi^{-3/2}), \quad
\Psi_c^{+}(\xi) = O(\xi^{-1/2}), 
\]
as $\RED{|\xi|} \to \infty$ within the lower and the upper half-planes, respectively. \green{The latter ensures the uniqueness of the solution of the Wiener--Hopf problem. However, in the present paper we solely focus on the derivation of the Weiner--Hopf equations, and do not formulate growth conditions for problems discussed in the next sections.}

\subsection{The lattice problem}

The geometry of the problem is shown in Figure~\ref{fig:half_plane}, right. 
Let the Dirichlet boundary condition be satisfied on \RED{the positive} half-line,
\begin{equation}
u(m,0) = 0, \quad m\geq0,
\end{equation}
and let the homogeneous discrete Helmholtz equation (\ref{eq:Helm_lat}) be satisfied in the rest of the domain. Then, write the total field as a sum of an incident plane wave and a scattered field:
\[
u(m,n) = u^{\rm in}(m,n) + u^{\rm sc}(m,n), \quad u^{\rm in}(m,n) = (\lx^{\rm in})^{-m} (\ly^{\rm in})^{-n},
\]  
were $\lx^{\rm in}$ and $\ly^{\rm in}$ are some solutions to the dispersion equation (\ref{eq:lattice_Disp_eq}) with $|\lx^{\rm in}|>1$, $|\ly^{\rm in}|<1$, and the scattered field satisfies the radiation condition in terms of the limiting absorption principle. 
Using a symmetry argument, one can see that the scattered field is symmetric with respect to the line $n=0$. Therefore, this problem can also be restricted to the upper half-space with the following mixed boundary conditions:
\[
\ptl_{(m,0)}[u^{\rm sc}] = 0\text{ \RED{ for } } m<0,\quad \text{\RED{and}} \quad u^{\rm sc}(m,0) = -(\lx^{\rm in})^{-m}\text{ \RED{ for } } m\geq 0,
\]
\teal{where the discrete derivative $\ptl_{(m,0)} $ is defined in~\eqref{eq:disc_derivative}}. In analogy with the continuous case, the condition for $m<0$ will be referred to as a Neumann boundary condition. Note that it appears naturally from the discrete Helmholtz equation just by taking into account the symmetry
\[
u^{\rm sc}(m,1) = u^{\rm sc}(m,-1).
\]


Upon introducing the auxiliary function $w(m,n)=\lx^{m}\ly^n$ with \(D_d(\lx,\ly)~\!\!=~\!\!0\), we can apply \RED{the lattice} Green's identity  (\ref{Green's_lattice}) in the upper half-space (using the contour shown in Figure~\ref{fig:latticeGreenhalfplane} and taking the limit $R\to \infty$) \RED{to} $w$ and $u^{\rm sc}(m,n)$, to \RED{obtain the functional equation} 
\begin{equation}
\label{eq: WH_discrete}
 \Psi_d^-(\lx) =K_d(\lx)\Psi_d^+(\lx)+F_d(\lx),
\end{equation}
where
\[
K_d(\lx) = \frac{1}{\Upsilon(\lx)}, \quad F_d(\lx) = \frac{1}{1 - \lx\left(\lx^{\rm in}\right)^{-1}},
\]
and
\[
\Psi_d^+(\lx) = \sum_{m=0}^{\infty}\lx^{m}\ptl_{(m,0)}[u^{\rm sc}],\quad \Psi_d^-(\lx) = \sum_{m=-\infty}^{-1}\lx^{m}u^{\rm sc}(m,0).
\]
\RED{B}y definition $\Psi^+_d(\lx)$ and $\Psi^-_d(\lx)$ are analytic inside and outside the unit circle respectively\footnote{Hereafter the superscript ${}^+$ \RED{(resp.\ ${}^-$)} is used to denote a spectral function analytic in the upper \RED{(resp.\ lower)} half-space \RED{in the continuous case} or inside \RED{(resp.\ outside)} the unit circle \RED{in the discrete case}.
\RED{The} subscripts ${}_c$ and ${}_d$ are used to distinguish between continuous and discrete problems.}\RED{, and} 
\[
\Psi_d^-(\lx) = O(s^{-1}), \quad \text{as } \RED{|\lx|}\to\infty.
\]
The equation  (\ref{eq: WH_discrete}) is therefore a Wiener--Hopf problem on the unit circle with unknowns $\Psi^+_d$ and $\Psi^-_d$\RED{, kernel $K_d$ and forcing $F_d$}.

\section{The method of analogies}
\label{sec:analogy}
The equations (\ref{eq: WH_discrete}) and (\ref{eq: Wh_half_plane}) have a similar structure. Indeed, $\Psi_d^+(\lx),\Psi_d^-(\lx)$ are one-sided  (discrete) Fourier transforms (DFT), $\Psi^+_c(\xi), \Psi^-_c(\xi)$ are one-sided (integral) Fourier transforms (IFT), and both pairs are analytic in complementary domains \RED{of some} complex plane. Moreover, the functions $\Upsilon(\lx), i\gamma(\xi)$ are square roots that appear when differentiating 
(in the discrete case by differentiation operator we understand the finite difference operator $\ptl_{\bnu}$)  the outgoing waves $w(m,n)$ and $w(x,y)$ with respect to $m$ and $y$ respectively, on the line $m=0$ and $y=0$ respectively. \green{The quantities  $\Upsilon(\lx)$ and $i\gamma(\xi)$ will be referred to as propagation functions, as both are related to the vertical wavenumber of an outgoing wave.} 
Finally, the forcing in both problems is given by polar terms corresponding to the incident plane waves.

These similarities stem from the similarit\RED{ies between} the continuous and discrete Green's \RED{identities} 
 \eqref{eq:Green_cont_first} and \eqref{Green's_lattice}, and hold for other examples. \RED{The central point of the present article is to formalise these similarities through the following table of notations and  analogy.}
\begin{center}
\begin{tabular}{ | l | l| l | } 
 \hline
 {} &discrete setting &continuous setting \\
 \hline
 \RED{normal} derivative & $\ptl_{\bnu}$ & $\ptl/\ptl\bn$  \\
 \hline
 transforms & DFT & IFT\\
 \hline
  Wiener--Hopf contour & unit circle  & real axis \\
  \hline
  dispersion surface topology & torus & sphere\\
 \hline
 spectral variable & $\lx$ & $\xi$ \\
  \hline
  \green{propagation function} & $\Upsilon(\lx)$ & $i\gamma(\xi)$ \\
  \hline
   \green{incidence parameters} & \green{$\lx^{\rm in}$, $\Upsilon^{\rm in}$, $\ly^{\rm in}$} & \green{$\kstar$, $i\gamma^{\rm in}$}\\ 
   \hline
 plane wave & $\lx^{m} \left(\ly(s)\right)^n$& $\exp\{i\xi x  + i\gamma(\xi)y\}$\\
 \hline
a horizontal shift& $s^{M}$ & $\exp\{i\xi a\}$ \\
 \hline
a vertical shift& $q^{N}$ & $\exp\{i\gamma(\xi)b\}$ \\
 \hline
\end{tabular}
\end{center}
The vertical and horizontal shifts in the table above, refer to possible shifts in the boundary conditions, see Section~\ref{sec:stag} for example. \green{Additionally, for brevity, we have introduced the following notation:
\[
\Upsilon^{\rm in} \equiv \Upsilon(\ly^{\rm in}),\quad \gamma^{\rm in} \equiv \gamma(\xi^{\rm in})= k\sin\theta^{\rm in}. 
\]}

\begin{mdframed}
 \begin{analogy}
There is an analogy between the Wiener--Hopf equations for continuous and lattice problems: 
\[
\Psi^-_d(s) = K_d(s)\Psi^+_d(s) + F_d(s) \Longleftrightarrow \Psi^-_c(\xi) = K_c(\xi)\Psi^+_c(\xi) + F_c(\xi),
\]
 where
 \begin{itemize}
 \item the functions $\Psi^+_d(s)$ and $\Psi^-_d(s)$ are analytic inside and outside the unit circle, respectively and the Wiener--Hopf equation is valid on the unit circle.
     \item the functions ${\Psi}_c^-(\xi)$ and ${\Psi}_c^+(\xi)$ are analytical in the lower and the upper half-planes respectively and the Wiener--Hopf equation is valid on the line $(-\infty,\infty)$;
     \item  there is a function $\mathcal{K}\!$, called the \emph{generating kernel function}, such that
     \[
K_d(s) = \mathcal{K}\!\left[\Upsilon(\lx),\lx^M,\ly^N\right],\quad K_c(\xi) = \mathcal{K}\!\left[i\gamma(\xi),\exp\{i\xi a\},\exp\{i\gamma(\xi) b\}\right];
\]
     \item there is a function ${\mathcal{F}}\!$, called the \emph{generating forcing function}, such that
\begin{align*}
F_d(s) &=\mathcal{F}\!\left[1 - \lx\left(\lx^{\rm in}\right)^{-1},1 - \lx\lx^{\rm in},(\lx^{\rm in})^M,(\ly^{\rm in})^N,\Upsilon(s),\Upsilon(s^{\rm in})\right],\\
F_c(\xi) &=\mathcal{F}\!\left[i\kstar-i\xi,-i\kstar-i\xi,\exp\{i \xi^{\rm in} a\},\exp\{i\gamma(\xi^{\rm in}) b\},i\gamma(\xi),i\gamma(\xi^{\rm in})\right].
\end{align*}
 \end{itemize}
 \end{analogy}
 \end{mdframed}

\RED{Note that the number of variables that the generating functions depend upon might change from one problem to another, but their definition should be clear from the context. In the analogy above, we gave a general form that encompasses \green{most}\footnote{\green{In Section~\ref{subsec:waveguide_strip}, the situation differs slightly, since the forcing terms originate from boundary contributions associated with the $y$- and $n$-axes, respectively.}} of the examples presented in the next section. Moreover, in certain cases, we will obtain matrix Wiener-Hopf equations, in which case the kernel generating function will be a square matrix function and the forcing generating function will be a vector function.}

For the half-plane problem \RED{considered in Section \ref{sec: half_plane}} , \RED{the} generating functions can easily be written down explicitly:
\RED{
\begin{alignat*}{4}
 & K_d(s)=\mathcal{K}\!\left[ \Upsilon(s)\right], \quad && K_c(\xi)=\mathcal{K}\!\left[ i\gamma(\xi)\right], \quad && \text{with} \quad && \mathcal{K}\!\left[t\right] = \frac{1}{t} \\
 & F_d(s)=\mathcal{F}\!\left[ 1 - \lx\left(\lx^{\rm in}\right)^{-1} \right], \quad && F_c(\xi)=\mathcal{F}\!\left[ i\kstar-i\xi\right], \quad && \text{with} \quad && \mathcal{F}\!\left[t\right] = \frac{1}{t}
\end{alignat*}
}
\RED{In the next Section, several important two-dimensional} diffraction problems are treated similarly. \RED{The analogy is shown to hold in all cases} and the generating functions 
are \RED{given explicitly}. 


\section{Examples}
\teal{In this section, continuous and discrete formulation of diffraction problems will be considered side by side and the analogy formulated above will be demonstrated explicitly.}
\label{sec:examples}
\subsection{Diffraction by a Neumann half-plane}
Let us consider a problem of diffraction by a half-plane, but this time satisfying Neumann boundary conditions:
\[
\frac{\ptl u}{\ptl \RED{\boldsymbol{n}}}(x,0) = 0, \quad x>0, \quad \ptl_{(m,0)}[u]=0,\quad m\geq0
\]
The Wiener--Hopf equation for the continuous problem is of \RED{the} type (\ref{eq: Wh_half_plane}) with
\[
K_c(\xi) = i\gamma(\xi), \quad F_c = \frac{\gamma^{\rm in}}{\xi -\xi^{\rm in}},
\]
\[
\Psi_c^+(\xi)  = \int_{0}^{\infty} \exp\{i\xi x\}u^{\rm sc}(x,0) dx,\quad \Psi_c^-(\xi)  = \int_{-\infty}^{0} \exp\{i\xi x\}\frac{\ptl u^{\rm sc}}{\ptl y}(x,0) dx.
\]
The Wiener--Hopf equation for the lattice problem is of \RED{the} type (\ref{eq: WH_discrete}) with
\[
K_d(s) = \Upsilon(s), \quad F_d(s)= \frac{-\Upsilon^{\rm in}}{1-s\left(s^{\rm in}\right)^{-1}},
\]
\[
\Psi_d^-(\lx) = \sum_{m=-\infty}^{-1}\lx^{m}\ptl_{(m,0)}[u^{\rm sc}],\quad \Psi_d^+(\lx) = \sum_{m=0}^{\infty}\lx^{m}u^{\rm sc}(m,0).
\]

\RED{We therefore have
\begin{alignat*}{2}
K_c(\xi) &= \mathcal{K}\!\left[i\gamma(\xi)\right],&\quad& F_c(\xi) = \mathcal{F}\!\left[i\kstar-i\xi,i\gamma^{\rm in}\right], \\
K_d(s) &= \mathcal{K}\!\left[\Upsilon(s)\right],&\quad& F_d(\lx) = \mathcal{F}\!\left[1-s\left(s^{\rm in}\right)^{-1},\Upsilon^{\rm in}\right].
\end{alignat*}
for the generating functions $\mathcal{K}\!$ and $\mathcal{F}\!$ given by:
\[
\mathcal{K}\!\left[t\right] = t, \quad \mathcal{F}\!\left[t_1,t_2\right] = -\frac{t_2}{t_1} \cdot
\]}


\begin{rema}
By considering the continuous limit of the lattice problem, one \RED{finds} that there are multiple  choices for the discrete Neumann boundary conditions. For example, the following choice comes from the elastic interpretation of the lattice \cite{slepyan2012models,Sharma2015}:
\begin{equation}
\label{eq:Slepian_cond}
u(m+1,0)+u(m-1,0)+ u(m,1)+(\lk^2-3)u(m,0) = 0.  
\end{equation}
One can still use Green's identity to derive the Wiener--Hopf equation, but the analogy will change in this case.  Indeed, considering the boundary conditions (\ref{eq:Slepian_cond}) one can derive the following Wiener--Hopf equation:
\[
\RED{\bar \Psi}^-_d(s)   = \bar\Upsilon(s)\RED{\bar 
 \Psi}^+_d(s)+\frac{\bar \Upsilon(s^{\rm in})}{1-s\left(s^{\rm in}\right)^{-1}},
\]
where
\[
\bar \Psi_d^-(\lx) = \sum_{m=-\infty}^{-1}\lx^{m}\ptl_{(m,0)}[u^{\rm sc}]+\frac{1}{2}u(0,0)s^{-1},\quad \bar \Psi_d^+(\lx) = \sum_{m=0}^{\infty}\lx^{m}u^{\rm sc}(m,0),
\]
\[
\bar\Upsilon(s) = (s+s^{-1} + q(s) + (\lk^2-3)).
\]

Another way to establish the analogy in this case is to modify the definition of the lattice normal derivative while preserving the symmetry of the coefficients, which is necessary for Green's identity to hold.
\end{rema}
\subsection{Diffraction by a soft/hard half-plane}
Let us consider again a problem of diffraction by a half-plane, but let it satisfy Dirichlet boundary conditions on one side, and Neumann boundary conditions on the other side.  In the continuous case the boundary conditions are as follows: 
\[
u(x,0^+) = 0,\quad \frac{\ptl u}{\ptl \bn}(x,0^-)=0,\quad x>0,
\]
where the notation $0^\pm$ which corresponds to limits from above and below correspondingly. Generalising this notation to the lattice case we get
\[
u(m,0^+) = 0,\quad \ptl_{(m,0^-)}[u] = 0, \quad m\geq0,  
\]
where we suppose that \RED{on the} half-line $n=0,m>0$ all the nodes are duplicated and marked by $0^\pm$. It is worth mentioning that this problem in the continuous case was studied by many researchers using different methods~\cite{rawlins1975solution,hurd1976wiener,Daniele1978}.

Applying Green's identity twice in the upper and the lower half-planes we get the following matrix Wiener-Hopf equation:
\begin{equation}
\label{eq:WH_mat}
\rmU^-_c = \rmA_c\rmU_c^++\rmF_c, 
\end{equation}
where \footnote{\green{From now on,  we omit the variable dependence in the notation for the functions $\gamma(\xi)$, $\Upsilon(s)$ and $q(s)$ for brevity.}}
\[
\rmU^\pm_c = 
\begin{pmatrix}
W^\pm_c \\
U^\pm_c 
\end{pmatrix},\quad
\rmA_c =\frac{1}{2} 
\begin{pmatrix}
1 & i\gamma \\
-(i\gamma)^{-1} & 1 
\end{pmatrix},\quad 
\rmF_c =
\frac{1}{2(\xi - \xi^{\rm in})}
\begin{pmatrix}
\gamma + \gamma^{\rm in}
\\
i - i\gamma^{\rm in}\gamma^{-1}
\end{pmatrix},\]
\[
W^-_c = \int_{-\infty}^0 \exp\{i\xi x\}\frac{\ptl u^{\rm sc}}{\ptl y}(x,0)dx, \quad W^+_c = \int_0^\infty \exp\{i\xi x\}\frac{\ptl u^{\rm sc}}{\ptl y}(x,0^+)dx,
\]
\[
U^-_c = \int_{-\infty}^0 \exp\{i\xi x\}u^{\rm sc}(x,0)dx, \quad U^+_c = \int_0^\infty \exp\{i\xi x\}u^{\rm sc}(x,0^-)dx,
\]
for the continuous problem, and
\begin{equation}
\label{eq:WH_mat_lat}
\rmU^-_d = \rmA_d\rmU_d^++\rmF_d, 
\end{equation}
\[
\rmU^\pm_d = 
\begin{pmatrix}
W^\pm_d \\
U^\pm_d 
\end{pmatrix},\quad
\rmA_d =\frac{1}{2} 
\begin{pmatrix}
1 & \Upsilon \\
-\Upsilon^{-1} & 1 
\end{pmatrix},\quad 
\rmF_d =
\frac{-1}{2(1 - \lx\left(\lx^{\rm in}\right)^{-1})}
\begin{pmatrix}
\Upsilon + \Upsilon^{\rm in}\\
\Upsilon^{\rm in}\Upsilon^{-1} - 1 
\end{pmatrix},
\]
\[
W_d^-(\lx) = \sum_{m=-\infty}^{-1}\lx^{m}\ptl_{(m,0)}[u^{\rm sc}],\quad W_d^+(\lx) = \sum_{m=0}^{\infty}\lx^{m}\ptl_{(m,0^+)}[u^{\rm sc}],
\]
\[
U_d^-(\lx) = \sum_{m=-\infty}^{-1}\lx^{m}u^{\rm sc}(m,0),\quad U_d^+(\lx) = \sum_{m=0}^{\infty}\lx^{m}u^{\rm sc}(m,0^-).
\]
for the lattice problem. Note that \RED{for both the} continuous and lattice problems, the matrix kernel ${\rmA}_{d,c}$  \RED{belongs to} the \green{Chebotarev--Daniele--Khrapkov} class (also \RED{known as the} Daniele-Khrapkov class). However, the deviator polynomial (\green{the definition of the deviator polynomial is given, for example, in \cite{Rogosin2015}})  for \RED{the} continuous problem is of degree $2$, \RED{while} for \RED{the} lattice problem \RED{it} is of degree $4$, which makes \RED{the} lattice problem much harder to solve \RED{in practice}. It was shown in \cite{khrapkov1971some} that for \RED{a deviator} polynomial of degree $2$ the matrix \RED{Wiener--Hopf} problem can be reduced to a scalar one which \RED{in turn can be} solved using Cauchy's integral method. It was shown in \cite{zverovich1971boundary, antipov2002factorization} that for \RED{deviator} polynomials of higher degree, the solution of the related scalar problem has an exponential growth. In order to obtain a solution with polynomial growth\RED{,} one need\RED{s} to make an additional step\RED{:} solve the so-called Jacobi's inversion problem, which is a problem for a system of non-linear equations.  \RED{For} the special case of \RED{a} polynomial of degree $4$, it is reduced to the inversion of \RED{an} elliptic integral \cite{antipov1991exact}.  Another approach was used in \cite{Daniele1984}, leading to a different system of non-linear equations, which for a polynomial of degree $4$ \RED{also reduces} to the inversion of \RED{an} elliptic integral. 

\RED{The analogy holds since we have
\begin{alignat*}{2}
\rmA_c(\xi) &= \gK\left[i\gamma\right], &\quad & \rmF_c(\xi)=\gF\left[i\kstar - i\xi,i\gamma,i\gamma^{\rm in}\right]. \\
\rmA_d(\lx) &= \gK\left[\Upsilon\right], &\quad& \rmF_d(\lx)=\gF\left[1 - \lx\left(\lx^{\rm in}\right)^{-1},\Upsilon,\Upsilon^{\rm in}\right],
\end{alignat*}
for the generating kernel matrix function $\gK$ and forcing vector function $\gF$ given by:
\[
\gK[t] = \frac{1}{2} 
\begin{pmatrix}
1 & t \\
(-t)^{-1} & 1 
\end{pmatrix},
\quad 
\gF[t_1,t_2,t_3] =
-\frac{1}{2t_1}
\begin{pmatrix}
t_2 + t_3\\
t_3(t_2)^{-1}-1
\end{pmatrix}.
\]}

\subsection{\RED{Diffraction by a} right-angled wedge}
\green{Wedge problems have been studied intensively for over a century (see the review \cite{Nethercote2020}, for example). A Wiener--Hopf technique was initially employed by \cite{shanin1998excitation,Daniele2003-ib}, who introduced a pair of so-called generalized Wiener--Hopf equations, which were then reduced to a scalar Wiener--Hopf equation via a carefully chosen change of variables. In this work, we take a different approach by formulating a matrix Wiener--Hopf problem directly \cite{Daniele2014,Aitken2024,korolkov2024recycling}.}

Let us consider the problem of diffraction by a Dirichlet right-angled wedge and its lattice counterpart. 

\begin{figure}[htbp!]
    \centering
    \includegraphics[width=0.8\linewidth]{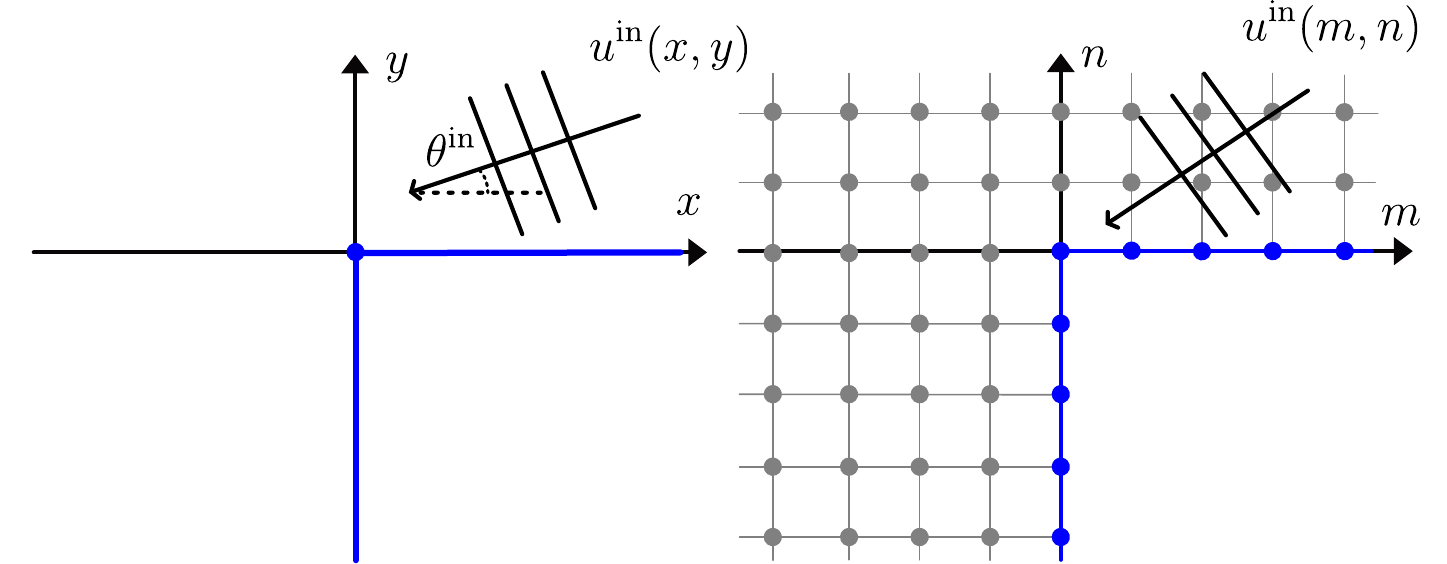}
    \caption{Geometry of the continuous \RED{(left)} and lattice \RED{(left)} problems of diffraction by a right-angled wedge}
    \label{fig:wedge_geom}
\end{figure}

The geometry is shown in Figure~\ref{fig:wedge_geom}. 
Using Green's identity (see \cite{korolkov2024recycling} for details) one can derive a \RED{$3 \times 3$ matrix Wiener--Hopf equation of the type (\ref{eq:WH_mat}) for the problem with}
\[
{\rmU}^-_c = 
\begin{pmatrix}
-W_c^+(-\xi)\\
W_c^-(\xi)\\
U_c^-(\xi)
\end{pmatrix},
\quad
{\rmU}^+_c =
\begin{pmatrix}
W^+_c(\xi)\\
-W^-_c(-\xi)\\
 -U^-_c(-\xi)
\end{pmatrix},
\]
\[
U^-_c (\xi)= \int^{0}_{-\infty} \exp\{i\xi x\} u^{\rm sc}(x,0) dx,
\]
\[
\quad W^-_c (\xi)= \int^{0}_{-\infty}\exp\{i\xi x\} \frac{\partial u^{\rm sc}}{\partial y}(x,0^-) dx,
\quad W^+_c (\xi)= \int_{0}^{\infty}\exp\{i\xi x\} \frac{\partial u^{\rm sc}}{\partial y}(x,0) dx.
\]
\[
{\rmA}_c = \frac{i}{2\gamma}
\begin{pmatrix}
0 &2i\gamma&2\gamma^2\\
i\gamma & i\gamma& -\gamma^2 \\
-1 & 1 & i\gamma 
\end{pmatrix},\quad
\renewcommand{\arraystretch}{2}
\rmF_c = -\gamma^{\rm in}
\begin{pmatrix}
\dfrac{2}{\xi+ \xi^{\rm in}}\\
\dfrac{1}{\xi- \xi^{\rm in}}\\
\dfrac{i}{(\xi- \xi^{\rm in})\gamma}
\end{pmatrix},
\]
for the continuous problem, and of the type (\ref{eq:WH_mat_lat}) with
\[
{\rmU}^-_d = 
\begin{pmatrix}
-W_d^+\left(\lx^{-1}\right)\\
W_d^-(\lx)\\
U_d^-(\lx)
\end{pmatrix},
\quad
{\rmU}^+_d =
\begin{pmatrix}
W^+_d(\lx)\\
-W^-_d\left(\lx^{-1}\right)\\
 -U^-_d\left(\lx^{-1}\right)
\end{pmatrix},
\]
\[
W_d^-(\lx) = -\sum_{m=-\infty}^{-1}\lx^{m}\ptl_{(m,0^-)}[u^{\rm sc}],\quad W_d^+(\lx) = \sum_{m=0}^{\infty}\lx^{m}\ptl_{(m,0)}[u^{\rm sc}],
\]
\[
U_d^-(\lx) = \sum_{m=-\infty}^{-1}\lx^{m}u^{\rm sc}(m,0),
\]
\[
{\rmA}_d = \frac{-1}{2\Upsilon}
\begin{pmatrix}
0 &2\Upsilon&-2\Upsilon^2\\
\Upsilon & \Upsilon& \Upsilon^2 \\
-1 & 1 & \Upsilon 
\end{pmatrix},\quad
\renewcommand{\arraystretch}{2}
\rmF_d = \Upsilon^{\rm in}
\begin{pmatrix}
\dfrac{2}{1-ss^{\rm in}}\\
\dfrac{1}{1-s\left(s^{\rm in}\right)^{-1}}\\
\dfrac{-1}{(1-s\left(s^{\rm in}\right)^{-1})\Upsilon}
\end{pmatrix},
\]
for the lattice problem. Here we \RED{assume} that the angle of incidence is restricted to $[0,\pi/2]$
and \RED{that the} total field is presented as a sum of scattered and geometric field in the upper half-space, and to be equal to the scattered field in the lower half-space, i.e.\ the scattered field has a jump of normal derivative \RED{across the} half-line $x<0,y=0$ (details are given in~\cite{korolkov2024recycling}). \green{Note that the definition of $u^{\rm sc}$ in this section is not the standard one used in the rest of the article. Indeed, it is not the total field minus the incident wave, but rather the total field minus the geometrical optics components. However, we prefer to keep it this way in order to be consistent with~\cite{korolkov2024recycling}. Note, that the definition of the scattered field does not affect the kernel of the Wiener--Hopf equation, only its forcing term.} 


Using the approach introduced in \cite{Daniele2014,Aitken2024} one can reduce both problems to \RED{a} $2\times 2$  matrix Wiener--Hopf equation with matrix kernel of Chebotarev--Daniele--Khrapkov class. Similarly to the previous section this kernel has a deviator polynomial of degree $2$ for the continuous problem, and a polynomial of degree $4$ for the lattice problem. It \RED{is} worth mentioning that \RED{the} lattice problem was solved analytically in \cite{Shanin2020,Shanin2022} using \RED{a} generalisation of the Sommerfeld integral method. To construct an explicit expression for the Sommerfeld transformant, the authors proposed an algorithm for constructing meromorphic functions on Riemann surfaces of genus 1. Alternatively, \RED{the} Wiener--Hopf approach puts this problem in the context of matrix factorisation, where it can be solved in a systematic way.    

\RED{Once again, the analogy holds since we have
\begin{alignat*}{2}
   \rmA_c(\xi) &= \gK\left[i\gamma\right], &\quad& \rmF_c(\xi)=\gF\left[i\kstar - i\xi,-i\kstar - i\xi, i\gamma,i\gamma^{\rm in}\right], \\
   \rmA_d(\lx) &= \gK\left[\Upsilon\right], &\quad& \rmF_d(\lx)=\gF\left[1-s\left(s^{\rm in}\right)^{-1},1-ss^{\rm in},\Upsilon,\Upsilon^{\rm in}\right],
\end{alignat*}
for the generating functions $\gK$ and $\gF$ given by
\[
\gK[t] = \green{-}\frac{1}{2t} 
\begin{pmatrix}
0 & 2t & -2t^2 \\
t &  t &  t^2 \\
-1 & 1 & t
\end{pmatrix},
\quad 
\gF[t_1,t_2,t_3,t_4] =
t_4
\begin{pmatrix}
2(t_2)^{-1}\\
(t_1)^{-1}\\
-(t_1t_3)^{-1}
\end{pmatrix}.
\]
}

\subsection{Finite strip}
\begin{figure}[htbp!]
    \centering
    \includegraphics[width=0.8\linewidth]{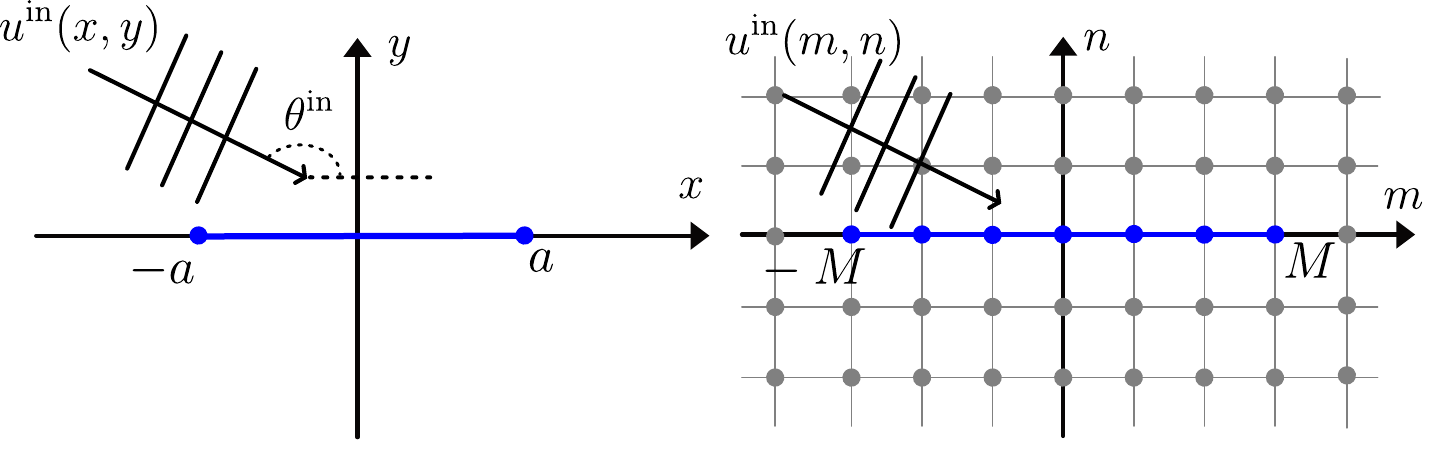}
    \caption{Geometry of the continuous and lattice problems of diffraction by a finite strip}
    \label{fig:strip_geometry}
\end{figure}

The geometry of the problem is shown in Figure~\ref{fig:strip_geometry}.

Let the Dirichlet boundary conditions be satisfied on the strip:
\[
u(x,0)=0,\quad x\in(-a,a),\quad u(m,0) = 0,\quad  m \in[-M,M].
\]

Using Green's identity \cite{korolkov2024recycling,kisil2018aerodynamic}  one can derive \RED{a} $2 \times 2$ matrix Wiener--Hopf equation of the type (\ref{eq:WH_mat}) for the continuous problem and of the type (\ref{eq:WH_mat_lat}) for the lattice problem with
\[
\rmU^\pm_c=
\begin{pmatrix}
U^\pm_c \\
W^\pm_c
\end{pmatrix},
\quad
{\rmA}_c=
\begin{pmatrix}
-\exp\{2i\xi a\} &  (i\gamma)^{-1} \\
0 &\exp\{-2i\xi a\}
\end{pmatrix},\quad
{\rm F}_c = \frac{i \exp\{i\kstar a\}}{\xi-\kstar}
\begin{pmatrix}
1\\
0
\end{pmatrix},
\]
\begin{align*}
U^-_c(\xi)  &= \exp\{i\xi a\}\int_{-\infty}^{-a} \exp\{i\xi x\}u^{\rm sc}(x,0) dx,\\
W^-_c(\xi)  &= \exp\{-i\xi a\}\int_{-a}^{a} \exp\{i\xi x\}\frac{\ptl u^{\rm sc}}{\ptl y}(x,0) dx,\\
U^+_c(\xi) &= \exp\{-i\xi a\}\int_a^\infty \exp\{i\xi x\}u^{\rm sc}(x,0)dx + \frac{i\exp\{-i\kstar a\}}{\xi-\kstar},\\
W^+_c(\xi)  &= \exp\{i\xi a\}\int_{-a}^{a} \exp\{i\xi x\}\frac{\ptl u^{\rm sc}}{\ptl y}(x,0) dx,
\end{align*}
for the continuous problem, and 
\[
\rmU^\pm_d=
\begin{pmatrix}
U^\pm_d \\
W^\pm_d
\end{pmatrix},
\quad
{\rmA}_d=
\begin{pmatrix}
-\lx^{2M} &  \Upsilon^{-1} \\
0 &\lx^{-2M}
\end{pmatrix},\quad
{\rm F}_d =   \frac{(\lx^{\rm in})^{M}}{1 - \lx\left(\lx^{\rm in}\right)^{-1}}
\begin{pmatrix}
1\\
0
\end{pmatrix},
\]

\begin{alignat*}{2}
U_d^-(\lx) &= \lx^{M}\sum_{m=-\infty}^{-M-1}\lx^{m}u^{\rm sc}(m,0),
& \quad &
U_d^+(\lx) = \lx^{-M}\sum_{m=M+1}^{\infty}\lx^{m}u^{\rm sc}(m,0)+\frac{(s^{\rm in})^{-M}}{1 - \lx\left(\lx^{\rm in}\right)^{-1}},\\
W_d^-(\lx) &= \lx^{-M}\sum_{m=-M}^{M}\lx^{m}\ptl_{(m,0)}[u^{\rm sc}],&\quad& W_d^+(\lx) = \lx^{M}\sum_{m=-M}^{M}\lx^{m}\ptl_{(m,0)}[u^{\rm sc}],
\end{alignat*}
\RED{for the lattice problem}. 
For both problems $\rmA_{c,d}$ is an upper triangular matrix, but for the continuous problem it grows exponentially at infinity, while for the lattice problem, it has polynomial growth. It is worth mentioning that the lattice problem was studied in \cite{sharma2015near2,medvedeva2024diffraction}. Another interesting observation is that kernel $\rmA_d$ falls into a class of branch commutative matrices \cite{shanin2012criteria}, i.e.\ it can be factorised using ideas \RED{from} \cite{zverovich1971boundary, antipov2002factorization}, and the solution of the problem can be presented in terms of elliptic integrals. Such a solution can be of practical importance, as there is no analytical solution available for the continuous problem. For example, one can study how the lattice solution approaches the solution of the continuous problem in the limit $h\to0$. 

\RED{The analogy holds in this case also since
\begin{alignat*}{2}
\rmA_c(\xi) &= \gK\left[i\gamma,\exp\{2i\xi a\}\right], & \quad & \rmF_c(\xi)=\gF\left[i\kstar - i\xi,\exp\{i\kstar a\}\right], \\
\rmA_d(\lx) &= \gK\left[\Upsilon,\lx^{2M}\right], &\quad& \rmF_d(\lx)=\gF\left[1 - \lx\left(\lx^{\rm in}\right)^{-1},(\lx^{\rm in})^{M}\right],
\end{alignat*}
where the generating functions $\gK$ and $\gF$ are given by:}
\[
\gK\left[t_1,t_2\right] = 
\begin{pmatrix}
t_2 & t_1 \\
0 & (t_2)^{-1} 
\end{pmatrix},
\quad 
\gF\left[t_1,t_2\right] =
\frac{t_2}{t_1}
\begin{pmatrix}
1\\
0
\end{pmatrix}.
\]

\subsection{Semi-infinite waveguide of two staggered plates}
\label{sec:stag}

\begin{figure}[htbp!]
    \centering
    \includegraphics[width=0.8\linewidth]{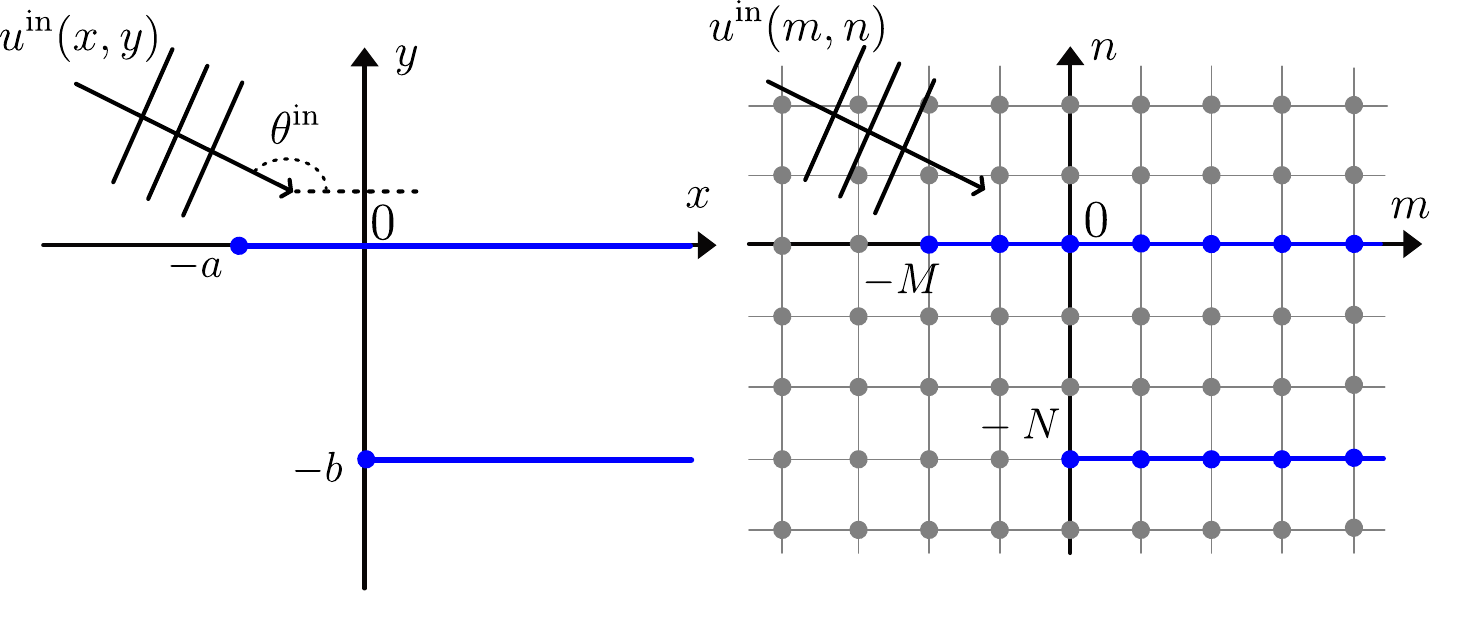}
    \caption{Geometries of continuous and lattice problems of diffraction by a half-infinite waveguide}
    \label{fig:stag_wave_geom}
\end{figure}

Let us consider \RED{the} problem of diffraction by a waveguide consisting of two staggered plates. The geometry is shown in Figure~\ref{fig:stag_wave_geom}, left. 
Let the total field satisfy Neumann boundary conditions on the \RED{plates}
. The problem was studied by many researchers \cite{vainshtein1969open, abrahams1988scattering,abrahams1990acoustic}, but there appears to be no derivation available  of the Wiener--Hopf equations using Green's identity. Thus, we are giving it below. 

\begin{figure}[htbp!]
    \centering
    \includegraphics[width=0.8\linewidth]{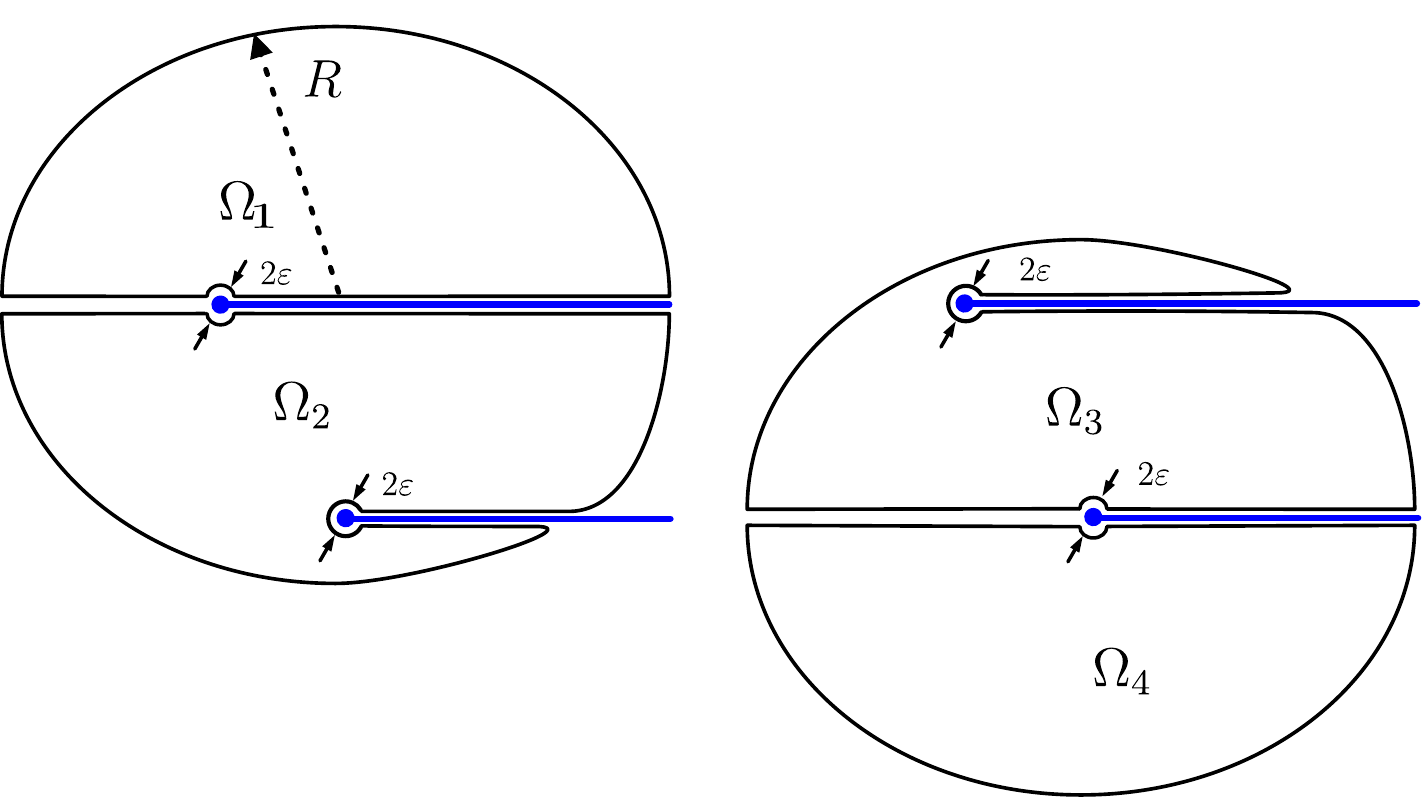}
    \caption{Domains for Green's identity for half-infinite waveguide problem}
    \label{fig:Greens_wave_domain}
\end{figure}

Consider the continuous problem. \RED{As a first step,} apply Green's identity \RED{to $u^{\rm sc}$ and the outgoing wave $w(x,y)$ defined in (\ref{eq:w_fun}) (and exponentially decaying in the upper-half space) on} the domain $\Omega_1$ shown in Figure~\ref{fig:Greens_wave_domain}, left. \RED{As before we consider a limiting procedure using some large arcs of radius $R$, but also some small arcs or radius $\varepsilon$ surrounding the edges of the plates. We then take the limit as $R\to \infty$ (possible due to the radiation condition) and $\varepsilon\to0$ (possible due to the Meixner conditions).}

\RED{This leads to the following functional equation}
\[
-i\gamma U^-_c + W^-_{c} - i\gamma U^+_{c^+} +W^+_{c}=0,
\]
where
\begin{alignat*}{2}
U^-_c(\xi)  &= \int_{-\infty}^{-a} \exp\{i\xi x\}u^{\rm sc}(x,0) dx,&\quad&
W^-_c(\xi)  = \int_{-\infty}^{-a} \exp\{i\xi x\}\frac{\ptl u^{\rm sc}}{\ptl y}(x,0) dx, \\
U^+_{c+}(\xi)  &= \int_{-a}^{\infty} \exp\{i\xi x\}u^{\rm sc}(x,0^+) dx,&\quad&
W^+_{c}(\xi)  = -\int_{-a}^{\infty} \exp\{i\xi x\}\frac{\ptl u^{\rm in}}{\ptl y}(x,0) dx.
\end{alignat*}
\RED{The second step is to apply} Green's identity \RED{on} the domain $\Omega_2$ shown in Figure~\ref{fig:Greens_wave_domain}, left \RED{but for the
functions $u^{\rm sc}$ and the wave $w(x,-y)$ (exponentially decaying in the lower-half space). This leads to}:
\[
i\gamma U^-_c + W^-_c  + i\gamma U^+_{c^-} + W^+_{c} - i\gamma\exp\{i\gamma h\}\left(\Phi^+_{c^+}- \Phi^+_{c^-}\right)=0,
\]
where
\[
U^+_{c-}(\xi)  = \int_{-a}^{\infty} \exp\{i\xi x\}u^{\rm sc}(x,0^-) dx,
\]
\[
\Phi^+_{c^+}(\xi)  = \int_{0}^{\infty} \exp\{i\xi x\}u^{\rm sc}(x,b^+) dx,\quad \Phi^+_{c^-}(\xi)  = \int_{0}^{\infty} \exp\{i\xi x\}u^{\rm sc}(x,b^-) dx.
\]
Adding the second functional equation to the first, we get 
\[
2W^-_c - i\gamma (U^+_c + \exp\{i\gamma b\}\Phi^+_c) = - 2W^+_c,
\]
where 
\[
\Phi^+_c = \Phi^+_{c^+}-\Phi^+_{c^-}, \quad  U^+_c = U^+_{c^+}-U^+_{c^-}.
\]
Repeating the same procedure, \RED{but} with the domains shown in Figure~\ref{fig:Greens_wave_domain}, right and \RED{the} shifted auxiliary function $w(x,y+b)$ we obtain the following equation:
\[
2\Xi^-_c - i\gamma(\Phi^+_c + \exp\{i\gamma b\}U^+_c) = -2 \Xi^+_c,
\]
where 
\[
\Xi^-_{c}(\xi)  = \int_{-\infty}^{0} \exp\{i\xi x\}\frac{\ptl u^{\rm sc}}{\ptl y}(x,-b)dx,\quad  \Xi^+_{c}(\xi)  = -\int_{0}^{\infty} \exp\{i\xi x\}\frac{\ptl u^{\rm in}}{\ptl y}(x,-b)dx .
\]
Finally, we arrive \RED{at a Wiener-Hopf} equation of the type (\ref{eq:WH_mat}) with 
\begin{align*}
\rmU^-_c &= 
\begin{pmatrix}
\exp\{i\xi a\}W^-_c \\
\Xi^-_c 
\end{pmatrix},\quad
\rmU^+_c =  
\begin{pmatrix}
\exp\{i\xi a\}U^+_c \\
\Phi^+_c 
\end{pmatrix}, \\
\rmF_c &=
\frac{\gamma^{\rm in}}{(\xi - \kstar)}
\begin{pmatrix}
\exp\{i\xi^{\rm in} a\}\\
\exp\{i\gamma^{\rm in} b\}
\end{pmatrix}, \quad 
\rmA_c =\frac{i\gamma}{2} 
\begin{pmatrix}
1 & \exp\{i\xi a + i\gamma b\} \\
\exp\{-i\xi a + i\gamma b\} & 1 
\end{pmatrix}.
\end{align*}
The latter is consistent with the equation derived in \cite{abrahams1988scattering}.

Similarly, \RED{by applying the discrete version of Green's identity, we obtain a} Wiener-Hopf equation of \RED{the} type (\ref{eq:WH_mat_lat}) for the lattice problem (the geometry is shown in Figure~\ref{fig:stag_wave_geom}, right) with
\begin{alignat*}{2}
\rmU^-_d &= 
\begin{pmatrix}
s^{M}W^-_d \\
\Xi^-_d
\end{pmatrix},&\quad
\rmU^+_d &=  
\begin{pmatrix}
s^{M}U^+_d \\
\Phi^+_d 
\end{pmatrix}, \\
\rmA_d &=\frac{\Upsilon}{2} 
\begin{pmatrix}
1 & \lx^{M}\ly^{N} \\
\lx^{-M}\ly^{N} & 1 
\end{pmatrix},
&\quad
\rmF_d &=
-\frac{\Upsilon^{\rm in}}{1 - \lx\left(\lx^{\rm in}\right)^{-1}}
\begin{pmatrix}
(\lx^{\rm in})^{M}\\
(\ly^{\rm in})^{N}
\end{pmatrix},
\end{alignat*}
\RED{where we have defined}
\begin{alignat*}{2}
U_d^+(\lx) &= \lx^{M}\sum_{m=-M}^{\infty}\lx^{m}\llbracket u^{\rm sc}(m,0) \rrbracket, &\quad \Phi_d^+(\lx) &= \sum_{m=0}^{\infty}\lx^{m}\llbracket u^{\rm sc}(m,-N)\rrbracket, \\
W_d^-(\lx) &= \lx^{M}\sum_{m=-\infty}^{-M-1}\lx^{m}\ptl_{(m,0)}[u^{\rm sc}],&\quad \Xi_d^-(\lx) &= \sum_{m=-\infty}^{-1}\lx^{m}\ptl_{(m,-N)}[u^{\rm sc}].
\end{alignat*}
\RED{In the above equation, by $\llbracket u^{\rm sc}\rrbracket$, we mean \green{the jump of $u^{\rm sc}$:
\[
\llbracket u^{\rm sc}(m,\cdot)\rrbracket = u^{\rm sc}(m,\cdot^+) - u^{\rm sc}(m,\cdot^-). 
\]
}
}
\RED{We should also note that a} closely related lattice problem with slightly different boundary conditions \RED{was} studied in \cite{maurya2019scattering}.

\RED{The analogy holds in this case too since we have}
\begin{align*}
\rmA_d(\lx) &= \gK\left[\Upsilon,\lx^{N},\ly^{M}\right],\quad  \rmA_c(\xi) = \gK\left[i\gamma,\exp\{i\xi a\},\exp\{i\gamma b\}\right], \\
\rmF_d(\lx)&=\gF\left[\Upsilon^{\rm in},1 - \lx\left(\lx^{\rm in}\right)^{-1},(\lx^{\rm in})^{M},(\ly^{\rm in})^{N}\right], \\
\rmF_c(\xi)&=\gF\left[i\gamma^{\rm in},i\kstar - i\xi,\exp\{i\kstar a\},\exp\{i\gamma(\kstar) b\}\right],
\end{align*}

\RED{where the generating functions $\gK$ and $\gF$ are given by:}
\[
\gK\left[t_1,t_2,t_3\right] = \frac{t_1}{2} 
\begin{pmatrix}
1& t_2t_3 \\
t_2(t_3)^{-1} & 1
\end{pmatrix},
\quad 
\gF\left[t_1,t_2,t_3,t_4\right] =
-\frac{t_1}{t_2}
\begin{pmatrix}
t_3\\
t_4
\end{pmatrix}.
\]

\subsection{Finite strip in a waveguide}
\label{subsec:waveguide_strip}
\begin{figure}[h]
    \centering
    \includegraphics[width=0.8\linewidth]{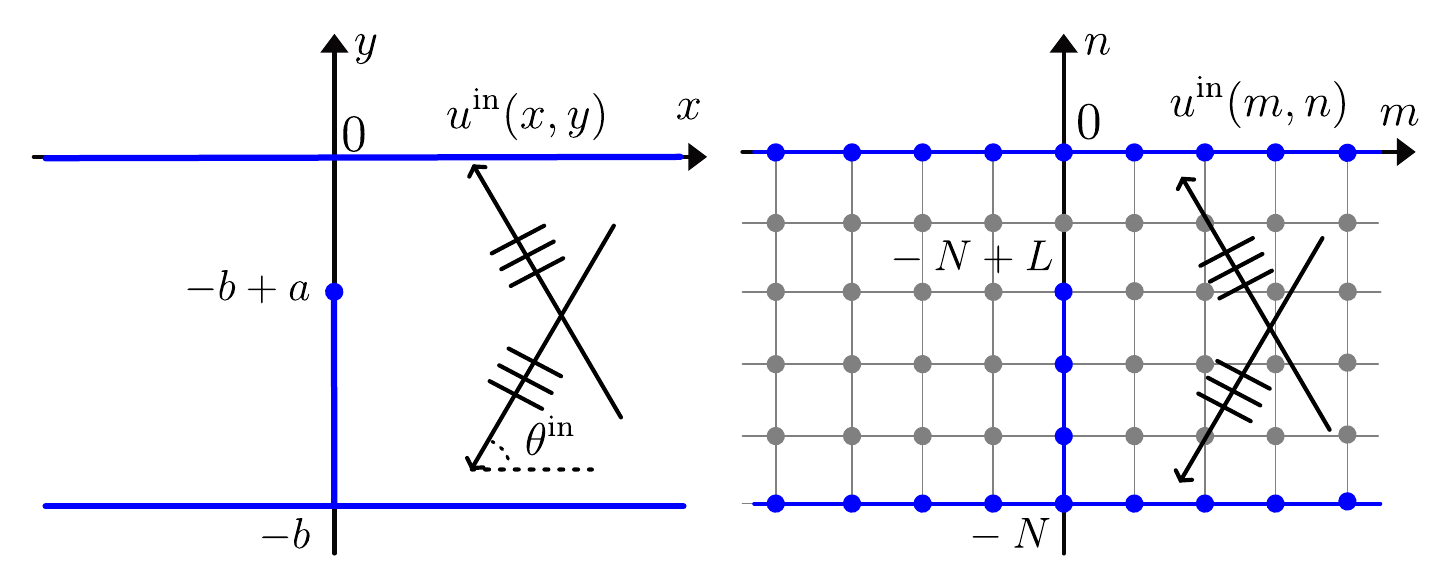}
    \caption{Geometries of continuous and lattice problems of diffraction by a strip in a waveguide}
    \label{fig:waveguide_barrier}
\end{figure}

Consider \RED{the} problem of diffraction by a strip in a waveguide, \RED{as} shown in Figure~\ref{fig:waveguide_barrier}, left. 
Let the total field satisfy Neumann boundary conditions on \RED{the} waveguide walls, and on the strip. The total field is presented as the sum of the incident and scattered fields:
\[
u = u^{\rm sc} + u^{\rm in},\quad u^{\rm in}  = \exp\{-ik\cos\theta^{\rm in}x\}\cos(k\sin\theta^{\rm in}y), 
\]
\RED{The incident angle} $\theta^{\rm in}$ is \RED{a} Brillouin partial angle of a guided wave \RED{defined for some $n\in \mathbb{Z}$ chosen such that \green{$|\pi n/kb|\leq 1$} by}
\[
\sin\theta^{\rm in} = \frac{\pi n}{k b}.
\]
Using \RED{a} symmetry argument, one can show that 
\[u^{\rm sc}(y,0)=0 \text{ \ for \ } y\in (-h+a,0).\]
\RED{Therefore,} only the right half of the waveguide needs to be considered.

\begin{figure}[h]
    \centering
    \includegraphics[width=0.4\linewidth]{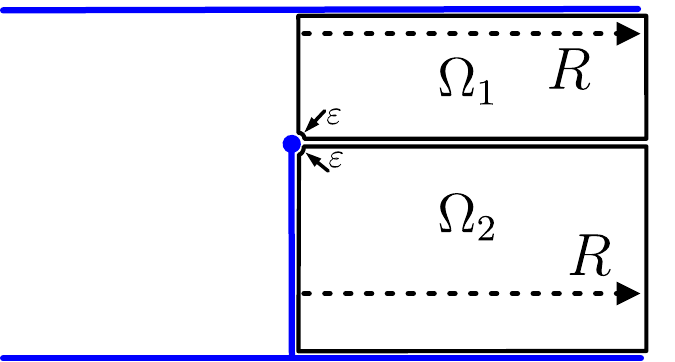}
    \caption{Green's domains for a waveguide with a strip}
    \label{fig:Greens_wg_barrier}
\end{figure}

Apply Green's identity in $\Omega_1$ (see Figure~\ref{fig:Greens_wg_barrier}) as it\RED{s} width $R$ tends to $\infty$  \RED{for the pairs $(u^{\rm sc},w_1)$ and $(u^{\rm sc},w_2)$,  the functions $w_{1,2}$ being defined by} 
\[
w_1 = w(x,-y)- w(-x,-y), \quad w_2 = w(x,y+b-a) -  w(-x,y+b-a).
\]
\RED{Do the same for $\Omega_2$ and the pairs $(u^{\rm sc},w_1)$ and $(u^{\rm sc},w_2)$, the functions $w_{3,4}$ being defined by}
\[
w_3 = w(x,-(y+b-a)) +  w(-x,-(y+b-a)), \quad w_4 = w(x,y+b) +  w(-x,y+b).
\]
\RED{This leads to} a system of 4 functional equations, which can be  reduced to \RED{a} $2\times 2$ Wiener--Hopf equation of the type (\ref{eq:WH_mat}) with:
\begin{align*}
\rmU^-_c &= 
\begin{pmatrix}
U^+_c(-\xi) \\
W^+_c(-\xi) 
\end{pmatrix},\quad
\rmU^+_c =  
\begin{pmatrix}
U^+_c(\xi) \\
W^+_c(\xi) 
\end{pmatrix}, \quad \rmF_c = \tilde f_c
\begin{pmatrix}
\cos(\gamma(b-a))\\
\gamma \sin(\gamma(b-a))
\end{pmatrix}, \\
\rmA_c &=\frac{-1}{\gamma\sin(\gamma b)} 
\begin{pmatrix}
\gamma \sin(\gamma (2a-b)) & 2\cos(\gamma a)\cos(\gamma (b-a)) \\
2\gamma^2\sin(\gamma a)\sin(\gamma (b-a)) & -\gamma \sin(\gamma (2a-b)) 
\end{pmatrix},
\end{align*}
\RED{where we defined} 
\[
\tilde f_c = 2i\kstar\frac{\gamma\cos(\gamma^{\rm in} (a-b))\sin(a\gamma) - \gamma^{\rm in}(\cos(\gamma a)\sin(\gamma^{\rm in}(a-b)) + \sin(\gamma^{\rm in} b))}{\gamma \sin(\gamma b)(\gamma -\gamma^{\rm in})(\gamma + \gamma^{\rm in})},
\]
\[
U^+_c(\xi)  = \int_{0}^{\infty} \exp\{i\xi x\}u^{\rm sc}(x,-b+a) dx,\quad
W^+_c(\xi)  = \int_{0}^{\infty} \exp\{i\xi x\}\frac{\ptl u^{\rm sc}}{\ptl y}(x,-b+a) dx.
\]
The latter is consistent with \cite{bookWH}.
Note that when $b =2a$ the kernel reduces to a matrix of the Chebotarev--Daniele--Khrapkov class with \RED{a} deviator polynomial of degree $2$. 

Let \RED{us now} consider the lattice problem with a strip of length $L$ and waveguide of width $N$ (the geometry is shown in Figure~\ref{fig:waveguide_barrier}, right).
\RED{In this case,} the incident wave \RED{takes} the following form:
\[
u^{\rm in}_{m,n} = (\lx^{\rm in})^{-m}((\ly^{\rm in})^n+(\ly^{\rm in})^{-n}),
\]
where $(s^{\rm in},q^{\rm in})$ is a point on the dispersion surface that corresponds to a discrete waveguide mode: 
\[
\ly^{\rm in} =\exp\{i\pi p/N\}, \quad p = 1,\ldots,N.
\]
\RED{A similar application of the discrete Green's identity, leads to a Wiener-Hopf equation of the type (\ref{eq:WH_mat_lat}) with:}
\begin{align*}
\rmU^-_d &= 
\begin{pmatrix}
U^+_d(1/s) \\
W^+_d(1/s) 
\end{pmatrix},\quad
\rmU^+_d =  
\begin{pmatrix}
U^+_d(s) \\
W^+_d(s) 
\end{pmatrix}, \quad \rmF_d = \tilde f_d
\begin{pmatrix}
(\ly^{N-L}+\ly^{-N+L})\\
-\Upsilon (\ly^{N-L}-\ly^{-N+L})
\end{pmatrix}, \\
\rmA_d &=\frac{1}{\Upsilon(\ly^N-\ly^{-N})} 
\begin{pmatrix}
-\Upsilon (\ly^{2L-N}-\ly^{-2L+N})  & (\ly^{L}+\ly^{-L})(\ly^{N-L}+\ly^{-N+L}) \\
\Upsilon^2(\ly^{L}-\ly^{-L})(\ly^{N-L}-\ly^{-N+L}) & \Upsilon (\ly^{2L-N}-\ly^{-2L+N})  
\end{pmatrix},
\end{align*}
\RED{where we have defined}
\begin{align*}
U_d^+(\lx) &= \sum_{m=1}^{\infty}\lx^{m}u^{\rm sc}(m,-N+L),\quad W_d^+(\lx) = \sum_{m=0}^{\infty}\lx^{m}\ptl_{(m,0)}[u^{\rm sc}], \text{ and }\\
\tilde f_d &= -\frac{\hat\Upsilon^{\rm in}\left((\ly^{\rm in})^{L-N} + (\ly^{\rm in})^{-L+N}\right)\left(\ly^L - \ly^{-L}\right)}{2\left(\ly^N - \ly^{-N}\right)\left(1-\ly(\ly^{\rm in})^{-1}\right)\left((1-\ly\ly^{\rm in})\right)}\\
&+\hat\Upsilon^{\rm in}\Upsilon^{\rm in}\frac{\left(\ly^L + \ly^{-L}\right)\left((\ly^{\rm in})^{L-N} - (\ly^{\rm in})^{-L+N}\right)+2\left((\ly^{\rm in})^{N} + (\ly^{\rm in})^{-N}\right)}{2\Upsilon\left(\ly^N - \ly^{-N}\right)\left(1-\ly(\ly^{\rm in})^{-1}\right)\left((1-\ly\ly^{\rm in})\right)},
\end{align*}
where $\hat\Upsilon^{\rm in} \equiv \Upsilon(\ly^{\rm in})$.
\RED{As in the continuous case}, when $N = 2L$, the kernel reduces to a matrix of the Chebotarev--Daniele--Khrapkov class, \RED{but, this time,} with \RED{a} deviator polynomial of degree $4$.

\RED{The analogy naturally holds for the kernel since}
\[
\rmA_d(\lx) = \gK\left[\Upsilon,\ly^{N},\ly^{L}\right],\quad  \rmA_c(\xi) = \gK\left[i\gamma,\exp\{i\gamma b\},\exp\{i\gamma a\}\right].
\] for the generating kernel matrix function $\gK$ given by:
\[
\gK\left[t_1,t_2,t_3\right] = \frac{1}{t_1(t_2 - (t_2)^{-1})}\times 
\]
\[
\begin{pmatrix}
-t_1(t_3^2(t_2)^{-1} - (t_3)^{-2}t_2)& (t_3 + (t_3)^{-1})(t_2(t_3)^{-1} - (t_2)^{-1}t_3) \\
(t_1)^2(t_3 - (t_3)^{-1})(t_2(t_3)^{-1} - (t_2)^{-1}t_3) & t_1(t_3^2(t_2)^{-1} - (t_3)^{-2}t_2)
\end{pmatrix}.
\]
We do not give the explicit expression for the forcing vector \RED{generating function} due to its unwieldiness. \green{Nevertheless, we indicate below the variables on which it depends:
\begin{align*}
\rmF_d(\lx)&=\gF\left[\Upsilon,\hat \Upsilon^{\rm in},\Upsilon^{\rm in},1-\ly(\ly^{\rm in})^{-1},1-\ly\ly^{\rm in}, q^L,q^N, \left(q^{\rm in}\right)^L,\left(q^{\rm in}\right)^N  \right],\\
\rmF_c(\xi)&=\gF\left[i\gamma, i\xi^{\rm in}, i\gamma^{\rm in},i\gamma^{\rm in} -i\gamma, -i\gamma^{\rm in}-i\gamma, \exp\{i\gamma a\},\exp\{i\gamma b\}, \exp\{i\gamma^{\rm in} a\},\exp\{i\gamma^{\rm in} b\}\right].
\end{align*}

It may seem surprising that to $i\xi^{\rm in}$  in the continuous case corresponds $\hat\Upsilon^{\rm in}$ in the discrete case. However, this follows from the differentiation rules with respect to the variables $x$ and $n$, respectively. Indeed, we have: 
\[
\frac{\ptl}{\ptl x}w(0,y) = i\xi w(0,y)  \text{ and }  \ptl_{(0,n)}[w] = \Upsilon(\ly)w(0,n). 
\]
}

\section{The method of analogies in 3D}
\label{sec:3D}
\subsection{Green's identity}
Green's identity can be easily extended to three-dimensional problems. Indeed, in the continuous case
we have
\[
\int_{\ptl \Omega}\left[u\frac{\ptl w}{\ptl \bn}-w\frac{\ptl u}{\ptl \bn}\right] d\RED{S}= \int_\Omega(fw - gu)dV,
\]
where  $w,u$  are solutions of \RED{the} 3D Helmholtz equations
\[
\Delta u(x,y,z) + k^2 u(x,y,z) = f(x,y,z),\quad \Delta w(x,y,z) + k^2 w(x,y,z) = g(x,y,z),  
\]
$\Omega$ is a closed domain with boundary $\ptl \Omega$, and $\bn$ is the unit inward normal on $\ptl \Omega$.  

To introduce  3D Green's identity on lattices we need, as in \RED{the} 2D case, to introduce an analogue of the normal derivative.  Let us consider a cubic lattice and a pair of functions that satisfy the discrete Helmholtz equation:
\begin{equation}
\Delta_{(m,n,\ell)}[u] + \lk^2u(m,n,\ell) = f(m,n,\ell),\quad
\Delta_{(m,n,\ell)}[w] + \lk^2w(m,n,\ell) = g(m,n,\ell),
\end{equation}
where $\Delta_{(m,n,k)}[u]$ is the 3D discrete Laplace operator, which is a 7-point finite difference approximation of the continuous Laplace operator:
\begin{align*}
\Delta_{(m,n,\ell)}[u] &= u(m+1,n,\ell)+u(m-1,n,\ell)+ u(m,n+1,\ell)+u(m,n-1,\ell) \\
&+u(m,n,\ell+1)+u(m,n,\ell-1) - 6u(m,n,\ell).
\end{align*}



\RED{As in the 2D case}, let $\Omega$ be a closed domain \RED{of a 3D lattice} with boundary $\ptl\Omega$, \RED{and introduce} the multi-index ${\bnu} = (m,n,\ell)$. We can \RED{then} re-write the Helmholtz equation\RED{s} in a \RED{condensed} form \RED{similar to (\ref{eq:disc_Helm})}: 
\begin{equation}
\label{eq:disc_Helm_3D}
\sum_{{\bmu}\in\Omega\cup\ptl\Omega}\beta_{{\bnu}{\bmu}}u_{{\bmu}} = f_{\bnu}, \quad \sum_{{\bmu}\in\Omega\cup\ptl\Omega}\beta_{{\bnu}{\bmu}}w_{{\bmu}} = g_{\bnu}, \quad {\bnu} \in \Omega,
\end{equation}
where the coefficients $\beta_{{\bnu}{\bmu}}$ are defined as follows:
\[
\beta_{{\bnu}{\bmu}}=\begin{cases}
1, & {\bnu}\neq {\bmu}, \text{ ${\bnu}$ is adjacent \RED{to} ${\bmu}$}\\
\lk^2 - 6, & {\bnu} = {\bmu},\\
0, & \text{otherwise}.
\end{cases}
\]

Finally, \RED{we} introduce the discrete normal derivative
\[
\ptl_{\bnu}[u] = \sum_{{\bmu} \in \ptl \Omega\cup\Omega_{\rm adj}} \alpha_{{\bnu}{\bmu}} u_{\bmu} ,\quad {\bnu}\in \ptl\Omega,
\]
where $\Omega_{\rm adj}$ is the set of all nodes of $\Omega$ adjacent to the boundary.
There are 7 possible boundary configurations,  $\alpha_{{\bnu}{\bmu}}$ having the following structure:
\[
\alpha_{{\bnu}{\bmu}}=\begin{cases}
p/4, & {\bnu}\neq {\bmu}, \text{ ${\bnu}$ is adjacent \RED{to} ${\bmu}$, ${\bmu} \in \ptl \Omega$  }\\
1, & {\bnu}\neq {\bmu}, \text{ ${\bnu}$ is adjacent \RED{to} ${\bmu}$, ${\bmu} \in \Omega_{\rm adj}$}\\
l(\lk^2 - 6)/8, & {\bnu} = {\bmu},\\
0, & \text{otherwise},
\end{cases}
\]
where $p$ is some natural number between $1$ and $4$, and $l$ is some natural number between $1$ and $8$. The exact expression can be deduced by expanding the FEM procedure described in Appendix~\ref{app:FEM} to \RED{3D}. Below we consider only the problem of diffraction by a quarter-plane where $\alpha_{{\bnu}{\bmu}}$ corresponding to the planar part of the boundary is used. In this particular case $p=2$, $l=4$, as \RED{can be shown} directly from the symmetry of the problem.

\RED{As in the 2D case,} using  the symmetry of coefficients, we obtain the \RED{3D discrete} Green's identity:
\begin{equation}
\label{Green's_lattice_3D}
\sum_{{\bnu} \in \ptl \Omega} \left(\ptl_{\bnu}[u] w_{\bnu}  - \ptl_{\bnu}[w] u_{\bnu}\right) =    \sum_{{\bnu} \in \Omega}\left(g_{\bnu} u_{\bnu}-f_{\bnu} w_{\bnu}\right).
\end{equation}

\subsection{Dispersion \RED{relation}}
Let us seek solutions to the homogeneous Helmholtz equation 
\begin{equation}
\label{eq:Helm3d}
(\Delta +k^2)u(x,y\RED{,z}) = 0,
\end{equation}
where $k$ is the wavenumber\RED{, of the form}
\[
u(x,y) = \exp\{i\xi_1 x  + i\xi_2y+  i\gamma z \},
\]
\RED{for some complex numbers} $\xi_1,\xi_2,\gamma$.  Substituting the latter into the Helmholtz equation \RED{and introducing $\bdxi\equiv(\xi_1,\xi_2)$,} we get \RED{the dispersion relation}: 
\[
\RED{D_c}(\bdxi,\gamma) = -\xi_1^2 - \xi_2^2 - \gamma^2 +k^2 = 0.
\]
 Solving the latter with respect to $\gamma$ we obtain:
\[
\gamma(\bdxi)  = \sqrt{k^2 - \xi_1^2-\xi_2^2}.
\]
Similarly, let us \RED{seek} solutions of \RED{the} homogeneous discrete Helmholtz equation
\begin{equation}
\label{eq:Helm_lat_3d}
\Delta_{(m,n,\ell)}[u] + \lk^2u(m,n,\ell) = 0,
\end{equation}
\RED{where $\lk^2$ is the wavenumber, of the form}:
\begin{equation}
\label{eq:lattice_plane_wave_3D}
u(m,n,\ell) = \lx_1^m\lx^n_2 \ly^\ell.
\end{equation}  
\RED{for some complex numbers} $\lx_1,\lx_2, \ly$. Substituting the latter  into the \RED{discrete} Helmholtz equation and introducing $\blx\equiv (\lx_1,\lx_2)$, we get the dispersion relation:
\begin{equation}
\label{eq:disp_rel3D_lat}
D_{\RED{d}}(\blx,\ly) = \lx_1+ \lx_1^{-1} + \lx_2 + \lx^{-1}_2+\ly + \ly^{-1} + \lk^2 -6 = 0.
\end{equation}
Solving the latter with the respect to $\ly$ we get
\[
\ly(\blx) = -\frac{\lk^2 - 6 + \lx_1 + \lx_1^{-1}+\lx_2 + \lx_2^{-1}}{2} 
\pm \frac{\sqrt{\left(\lk^2 - 6 + \lx_1 + \lx_1^{-1}+\lx_2 + \lx_2^{-1}\right)^2 - 4}}{2}.
\]
\RED{As in the 2D case, it is helpful to introduce the function $\Upsilon$ defined by}
\RED{\[
\Upsilon(\blx) = \frac{q(\blx)-q^{-1}(\blx)}{2}=\frac{\sqrt{\left(\lk^2 - 6 + \lx_1 + \lx_1^{-1}+\lx_2 + \lx_2^{-1}\right)^2 - 4}}{2}.
\]}

\subsection{Diffraction by a Dirichlet quarter-plane}
\RED{The important canonical} problem of diffraction by a quarter-plane is the simplest generalisation of Sommerfeld's half-plane problem into 3 dimensions. \RED{In the continuous case, as we will briefly recall below, it is known that it can be formulated as a scalar Wiener--Hopf problem in two complex variables. Despite the fact that no closed-form solution to this problem is yet known, recent progress have been made by exploiting this two-complex-variables functional equation  \cite{assier2019diffraction,AssierAbrahamsSIAP21,AssierShanin2021Vertex,assier2024contribution}. To the authors knowledge, however, the discrete counterpart of the quarter-plane problem has not yet been considered. We will show below that it can also be formulated as a two-complex-variables Wiener--Hopf problem, and that an analogy, akin to that presented in Section~\ref{sec:analogy}, also holds in this case.}

\subsubsection{\RED{The} continuous problem}
\begin{figure}[htbp!]
    \centering
    \includegraphics[width=0.8\linewidth]{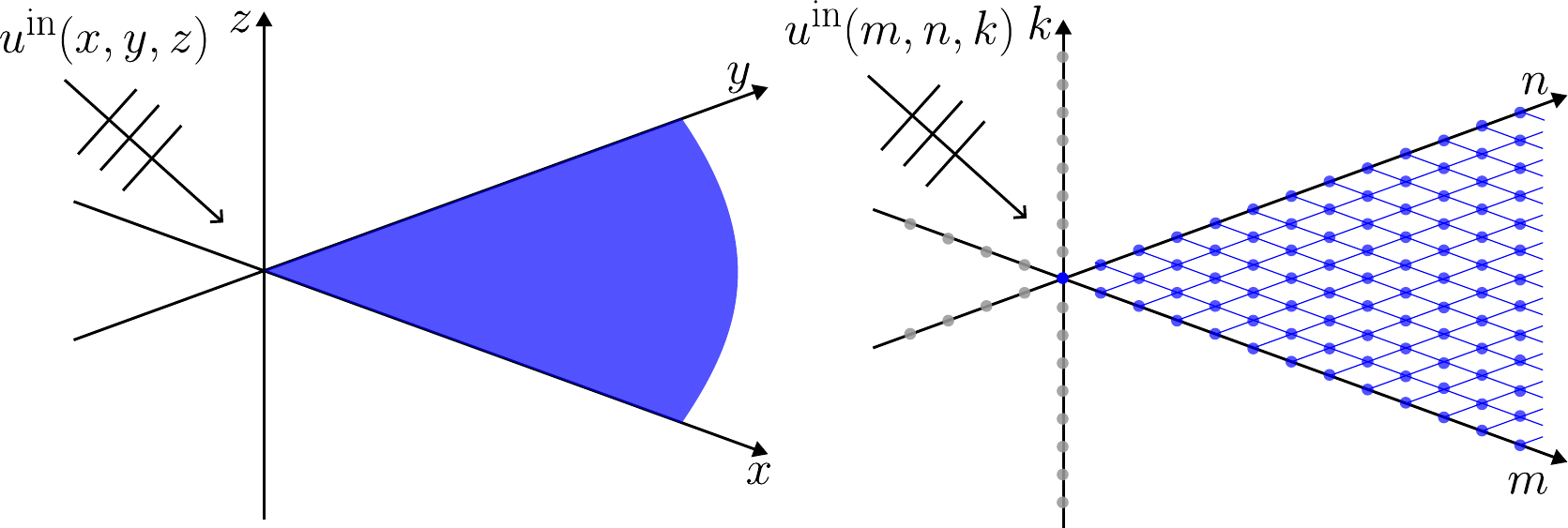}
    \caption{Geometry of the problem of diffraction by a quarter-plane}
    \label{fig:quarterplane}
\end{figure}

\RED{We study the diffraction of an incident plane wave $u^{\rm in}$ by a quarter-plane $\rm QP$ subjected to Dirichlet boundary conditions.} The geometry of the problem is sketched in Figure~\ref{fig:quarterplane}. 
The total field $u$ satisfies the Helmholtz equation 
\[
(\Delta + k^2)u(x,y,z) = 0
\]
everywhere except \RED{on} the quarter-plane, \RED{where a Dirichlet boundary condition is imposed:}
\[
\RED{u|_{\rm QP}=0, \quad {\rm QP}=\{x\geq 0,y\geq0,z=0\}}.
\]
The total field $u$ can be written as the sum of the incident plane wave $u^{\rm in}$ and the \RED{scattered} field $u^{\rm sc}$:
\[
u = u^{\rm sc} + u^{\rm in}, 
\]
\[
u^{\rm in}\RED{(x,y,z)} = \exp\{-i\xi^{\rm in}_1x - i\xi^{\rm in}_2y-i\xi^{\rm in}_3z\},\quad \xi^{\rm in}_3 = \sqrt{k^2 - \left(\xi^{\rm in}_1\right)^2 - \left(\xi^{\rm in}_2\right)^2}.
\]
\RED{For the problem to be well-posed,} the scattered field should satisfy the radiation condition as well as Meixner conditions at the edges and  at the vertex (see \cite{assier2019diffraction}). 

Using a symmetry argument, \RED{one} can show that $u^{\rm sc}$ satisfies Neumann boundary conditions on the rest of the plane $z=0$:
\[
\frac{\ptl u^{\rm sc}}{\ptl z}(x,y,0)=0, \quad \RED{(x,y,0)\notin {\rm QP}}.
\]

\RED{Following \cite{assier2019diffraction}, } the Wiener--Hopf equation \RED{can be derived by applying} Green's identity in the upper half-space $z>0$ (\RED{through} a limiting procedure similar to the one considered in 2D, see~Figure~\ref{fig:Green_dom_cont}, right), with the auxiliary  function
\[
w(\bdxi) = \exp\{i\xi_1 x +i\xi_2 y + i\gamma(\bdxi)z\}.
\]
After some algebra, we obtain: 
\begin{equation}
\label{eq:WH_3D}
U_c(\bdxi)=K_c(\bdxi)W_c(\bdxi) + F_c(\bdxi),  
\end{equation}
where \RED{the kernel $K_c$ and forcing $F_c$ are given by}
\[
K_c(\bdxi)=\frac{1}{i\gamma(\bdxi)},\quad F_c(\bdxi)=-\frac{1}{(\xi_1 - \xi^{\rm in}_1)(\xi_2 - \xi^{\rm in}_2)}.
\]
\RED{The unknown spectral function}  $W_c(\bdxi)$ is defined as
\[
W_c(\bdxi) = \iint_{\RED{{\rm Q}_1}} \frac{\ptl u^{\rm sc}}{\ptl z}(x,y,0\RED{^+})\exp\{i\xi_1 x + i\xi_2 y\}\green{dx dy},
\]
\RED{where ${\rm Q}_1=\{x\geq0,y\geq0\}$}, \RED{while} the unknown spectral function $U_c(\bdxi)$ \RED{is defined by}
\[
U_c(\bdxi) = \iint_{\mathbb{R}_2\RED{\setminus{\rm Q}_1}} u^{\rm sc}(x,y,0)\exp\{i\xi_1 x + i\xi_2 y\}\green{dx dy}.
\]
\RED{The} equation (\ref{eq:WH_3D}) is a two-complex-variables scalar Wiener--Hopf equation \RED{with kernel $K_c$, forcing $F_c$, and} unknowns $W_c$ and $U_c$. \RED{It is valid on the $(\xi_1,\xi_2)$ real plane}. To have a unique solution it should be supported with additional analyticity conditions, see \cite{assier2019diffraction}.

\subsubsection{\RED{The} lattice problem}
\RED{For the discrete quarter-plane problem, see Figure \ref{fig:quarterplane} (right),} the total field $u_{\bnu}$ satisfies the discrete Helmholtz equation
\[
\Delta_{\bnu}[u] + \lk^2 u_{\bnu} = \RED{0}
\]
everywhere except \RED{on} the quarter-plane, where \RED{it} satisfies Dirichlet boundary conditions:
\[
u_{\bnu} = 0, \quad {\bnu} \in {\rm QP} = \{m\geq 0, n\geq0,\ell =0\}. 
\]
The total field \RED{is the sum of the incident wave and the scattered field}: 
\[
u_{\bnu} = u^{\rm sc}_{\bnu} + u^{\rm in}_\bnu, \quad  u^{\rm in}(m,n,\ell) = (\lx_1^{\rm in})^{-m}(\lx_2^{\rm in})^{-n}(\ly^{\rm in})^{-\ell},
\]
where $\lx_1^{\rm in},\lx_2^{\rm in}, \ly^{\rm in}$ satisfy the dispersion relation (\ref{eq:disp_rel3D_lat}). The scattered field should satisfy the radiation condition, which as we mentioned before might not be possible for some values of $\lk$, which we exclude from \RED{our} consideration. 

Using a symmetry argument we can introduce a lattice Neumann boundary condition on the rest of the plane $k=0$: 
\[
\ptl_{(m,n,0)}[u^{\rm sc}] = 0,  \quad \RED{(m,n,0) \notin {\rm QP}},
\]
\RED{where} $\ptl_{(m,n,0)}[\cdot]$ is the lattice normal derivative \RED{defined by}
\begin{align*}
\ptl_{(m,n,0)}[u] &= \frac{1}{2}\left[u(m+1,n,0)+u(m-1,n,0)+ u(m,n+1,0)+u(m,n-1,0)\right] \\
&+(3-\lk^2/2)u(m,n,0)+u(m,n,1).
\end{align*}
Then, analogously to the continuous case, applying Green's identity in the upper half-space with the outgoing wave
\[
w(m,n,\ell) = \lx_1^m\lx_2^n\ly^\ell,
\]
we obtain the following functional equation:
\begin{equation}
\label{eq:WH_eq3D_lat}
U_d(\blx) = K_d(\blx)W_d(\blx) +  F_d(\blx). 
\end{equation}
\RED{The kernel $K_d$ and the forcing $F_d$ are given by}
\[
K_d(\blx)=\frac{1}{\Upsilon(\blx)}, \quad F_d(\blx) =  \frac{1}{\left(1 - \lx_1\left(\lx_1^{\rm in}\right)^{-1}\right)\left(1 - \lx_2\left(\lx_2^{\rm in}\right)^{-1}\right)},
\]
\RED{while the unknown spectral functions $U_d$ and $W_d$ are defined by}
\[
W_d(\blx) = \sum_{(m,n) \in \RED{{\rm Q}_1}} \ptl_{(m,n,0)}[u]\lx_1^m\lx_2^n, \quad U_d(\blx) = \sum_{(m,n) \in \RED{\mathbb{Z}^2\setminus{\rm Q}_1}} u(m,n,0)\lx_1^m\lx_2^n,
\]
\RED{where ${\rm Q}_1=\{(m,n)\in\mathbb{Z}^2: m\geq0, n\geq0\}$}.

\RED{The} equation (\ref{eq:WH_eq3D_lat}) is a 2D Wiener---Hopf equation \RED{valid on the} direct product of two unit circles (a torus). 

Equations (\ref{eq:WH_3D}) and (\ref{eq:WH_eq3D_lat}) have a similar structure, and the analogy can be built in the same way as it was done in Section~\ref{sec:analogy}. That is we can find some kernel generating function $\mathcal{K}$ and forcing generating function $\mathcal{F}$ such that
\RED{
\begin{alignat*}{2}
  & K_c(\bdxi)=\mathcal{K}\!\left[ i\gamma(\bdxi)\right]  \quad && F_c(\bdxi)=\mathcal{F}\!\left[ (i\xi_1 -i \xi_1^{\rm in})(i\xi_2 - i\xi_2^{\rm in})\right]\\
  & K_d(\blx)=\mathcal{K}\!\left[ \Upsilon(\blx)\right], \quad &&    F_d(\blx)=\mathcal{F}\!\left[ \left(1 - \lx_1\left(\lx_1^{\rm in}\right)^{-1}\right)\left(1 - \lx_2\left(\lx_2^{\rm in}\right)^{-1}\right) \right].
\end{alignat*}
For the quarter-plane problem we therefore recover the same generating functions as for the half-plane problem, namely
\[
\mathcal{K}\!\left[t\right]=\frac{1}{t}, \quad \mathcal{F}\!\left[t\right]=\frac{1}{t}.
\]
}


\section{Conclusions \RED{and perspectives}}\label{sec:conclusion}

\RED{In this article, we proposed an analogy between the Wiener-Hopf \textit{formulation} of scattering problems in the continuous setting and the Wiener-Hopf formulation of their discrete counterparts. This \teal{was} done through introducing a generalisation of Green's identity to the discrete setting (given in~\eqref{Green's_lattice}), which required an appropriate definition of a discrete normal derivative (given in~\eqref{eq:disc_derivative}). 

The main idea of the analogy (formulated in Section~\ref{sec:analogy}) is that the continuous kernel $K_c(\xi)$ and the discrete kernel $K_d(s)$ can be expressed in terms of the same kernel generating function $\mathcal{K}$. The same \teal{holds} for the continuous and discrete forcing terms $F_c(\xi)$ and $F_d(s)$,  \teal{which} can be expressed in terms of the same forcing generating function $\mathcal{F}$. This analogy, shown to hold for a very wide range of examples, holds  for \teal{scalar,} matrix and for two-complex-variables Wiener--Hopf problems.

One of our motivations was to show that if a Wiener--Hopf problem is already formulated for a continuous problem, then we can use the analogy to automatically formulate the Wiener-Hopf equation for its discrete counterpart. However, we did not focus here on the similitude between the \textit{resolution} of those Wiener--Hopf equations. We can however make the following remarks:}



\begin{itemize}
    \item  \RED{For a matrix Wiener--Hopf equation}, if either \(\rmA_c(\xi)\) or \(\rmA_d(s)\) \RED{belongs to the} Chebotarev--Daniele--Khrapkov class, then \RED{the same is true of the other matrix kernel. Therefore,}
    the analogy preserves this class. \RED{An important difference, however, is that
the degree of the deviator polynomial can differ between the continuous and the discrete case.} \RED{Indeed, we provided examples for which this degree is 2 for the continuous case and 4 for the discrete case.} \RED{This is significant, since the case of a degree 2 is known to be reducible}
to two scalar problems, while the degree 4 requires a more involved  procedure~\cite{antipov1991exact,Daniele1984} \RED{to factorise the kernel}.

\item \RED{A class that is however not preserved is that of rational kernels. Indeed, rational terms that may for instance arise due to some vertical shift in the discrete setting, might result in exponentially growing terms in the continuous setting. This is significant since the generalised Liouville theorem (a key result to solve Wiener--Hopf equations) can only be applied to functions with algebraic growth, not to those with exponential growth.}

\teal{\item For a class of discrete diffraction problems posed in a two dimensional waveguide, the kernel function will be a rational function with  finitely many poles. Such problems are directly solvable by simple procedures such as pole removal. In the continuous setting, however, there will be infinitely many poles in the kernel function, which would require more care in the construction of the solution.

\item In the discrete setting the kernel has an annulus of analyticity around the unit circle (with some possible deformations around branch points). This implies the possibility for fast evaluation of Cauchy type \green{integrals} and for approximating kernel functions  by a rational function with arbitrary precision (Runge's theorem). In the continuous case, if the kernel has a branch cut due to \(\gamma\) then the branch cut crosses the real line at infinity giving a jump of the kernel across it. This causes numerical difficulties in the evaluation of Cauchy type integrals and also makes it impossible to approximate the kernel with arbitrary accuracy on the whole real line by rational functions. }


\end{itemize}

\RED{We will now conclude by offering some perspectives and potential follow-ups to this work.} 

\begin{itemize}

\item \RED{An important concept when considering matrix kernel factorisation is that of partial indices~\cite{WHreviewAK}. An interesting question would be to investigate whether or not the analogy presented here preserves the partial indices.} 

\item Additionally, as hinted in introduction, \RED{we believe that a very similar} analogy can be established for the global relations of Fokas' \RED{unified transform} method, \RED{allowing one to translate} the methodology and results of Fokas' method to the discrete setting.

\item Finally, \RED{an important concept of diffraction theory is that of embedding formulae \cite{craster2003embedding}. These formulae can be used to simplify the calculation of wave fields directivities. They are generally valid for continuous 2D scattering problems, but can also be generalised to 3D \cite{shanin2005modified,skelton2010embedding,assier2012diffraction}. Interestingly, they have recently been shown to be linked to the Wiener--Hopf technique \cite{korolkov2024recycling}. Our analogy therefore suggests that one could also derive embedding formulae in the discrete setting.} 
\end{itemize}

\RED{We will investigate these ideas in our future work.}

\green{
\section*{Acknowledgment}
 A.V.K. is supported by a Royal Society
Dorothy Hodgkin Research Fellowship which also supported A.I.K via the Royal Society Research Fellows Enhanced Research Expenses.
Authors gratefully acknowledge the support of the EU H2020 grant MSCA-RISE-2020-101008140-EffectFact. The authors would also like to thank the Isaac Newton Institute for Mathematical Sciences (INI) for their support and hospitality during the programme ``WHT Follow on: the applications, generalisation and implementation of the Wiener--Hopf Method''(WHTW02), where work on this paper was undertaken and supported by EPSRC grant no EP/R014604/1. 
}



\bibliographystyle{unsrturl}
\bibliography{Bibliography}
\appendix
\appendixpage
\section{FEM derivation of lattice equations}
\label{app:FEM}
\begin{figure}
    \centering
    \includegraphics[width=0.7\linewidth]{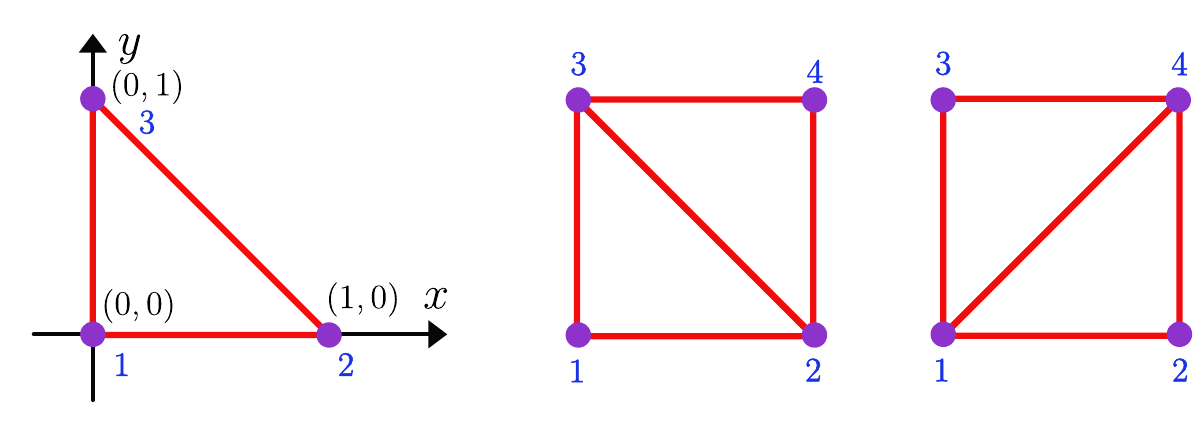}
    \caption{Unit triangle as a finite element of a square lattice and two distinct partitions of a square into a sum of two right-angled triangles}
    \label{fig:triangle}
\end{figure}

Consider  a weak \RED{formulation}  for the Helmholtz equation (\ref{eq:Helm}) in some finite domain $\Omega$:
\[
\int_\Omega \phi \Delta u \, dS  + k^2\int_\Omega \phi u \, dS = 0,  
\]
where $\phi$ is a test function from some appropriate function space. Applying the first Green's identity to the first term in the equation above we obtain: 
\[
\int_\Omega \phi \Delta u \, d\RED{S}  = -\int_{\Omega}\nabla \phi \cdot \nabla u \, d\RED{S}  - \int_{\ptl \Omega}\phi\frac{\ptl u}{\ptl n}dl. 
\]
\RED{Assuming that $u$ satisfies}
Neumann boundary conditions on $\ptl \Omega$, \RED{the boundary integral vanishes and we obtain}
\[
-\int_\Omega \nabla \phi \cdot \nabla u \, dS  + k^2\int_\Omega \phi u \, dS = 0.
\]
Following the ideology of the finite element method \cite{Zienkiewicz2013-pm}, \RED{we discretise the domain $\Omega$ into a regular square grid and number the resulting nodes. Each square element is seen as composed of two triangular elements.} \RED{If the total number of nodes is $L$}, \RED{we} approximate the field as follows:
\[
u(x,y) = \sum_{{\nu} =1}^L u_{\nu} N_{{\nu}}(x,y), 
\]
where $u_{\nu}$ are the nodal values of the field,
and $N_{\nu}(x,y)$ are linear shape functions. The shape function $N_{\nu}(x,y)$ is defined in such a way that it is 1 on the node ${\nu}$ and zero on all other nodes. It is convenient for calculations to define 3 basis shape functions over a unit triangle (see Figure~\ref{fig:triangle}, left):
\[
N_1(x,y) = 1-x-y, \quad  N_2(x,y) = x, \quad N_3(x,y) = y.
\]
To finish the derivation of the finite element equations, one needs to choose \RED{some} test functions $\phi$. Following Galerkin's method, we choose
\[
\phi = N_{\nu},\quad {\nu} = 1\ldots L.
\]
Thus, we obtain the following system of linear equations:
\[
{\rm K} {\rm u} + k^2{\rm M} {\rm u} = 0, 
\]
where \RED{${\rm u}$ is the column vector of the nodal values $u_\nu$.} ${\rm K}$ and ${\rm M}$ are the so-called stiffness and mass matrices \RED{defined by}
\begin{equation}
\label{eq:FEM_matrices}
{\rm K}_{ij} = -\int_\Omega \nabla N_i \cdot \nabla N_j dS, \quad {\rm M}_{ij} = \int_\Omega N_i N_j dS.
\end{equation}
 
A standard approach to evaluate ${\rm K}$ and ${\rm M}$ is to use the assembling procedure. Namely, those matrices are calculated once for a unit triangle, and then the ``global" matrix is calculated by adding multiple ``shifted'' element matrices. More formally, let us present the integrals (\ref{eq:FEM_matrices}) as sums over finite elements (triangles):
\[
{\rm K}_{ij} = \sum_{n} {\rm K}^{(n)}_{ij},\quad {\rm M}_{ij} = \sum_{n} {\rm M}^{(n)}_{ij},
\]
\[
{\rm K}^{(n)}_{ij} = -\int_{\Omega^{(n)}} \nabla N_i \cdot \nabla N_j dS, \quad {\rm M}^{(n)}_{ij} = \int_{\Omega^{(n)}} N_i N_j dS,
\]
where $n$ is the index that goes through all finite elements, and $\Omega^{(n)}$ is \RED{the} part of space corresponding to the n-th element. Due to the locality of shape functions, matrices  
${\rm K}^{(n)}$ and ${\rm M}^{(n)}$ have nonzero elements only at the intersections of rows and columns whose indices correspond to the indices of nodes belonging to the \RED{$n$th} element. Thus, instead of the matrices ${\rm K}^{(n)}$ or ${\rm M}^{(n)}$ it makes sense to consider corresponding non-zero submatrices ${\rm \tilde K}^{(n)}$ and ${\rm \tilde M}^{(n)}$, which are of size $3\times 3$ for triangular finite elements. Moreover, in the case of regular mesh\RED{, the} submatrices ${\rm \tilde K}^{(n)}$ or ${\rm \tilde M}^{(n)}$ are equal up to a permutation of columns and rows, so they \RED{only need to be calculated} once.

Let us consider \RED{the} triangle shown in Figure~\ref{fig:triangle}, left. Denote the element matrices for this triangle as ${\tilde{\rm K}}^e$ and $\tilde{\rm M}^e$. After some algebra, get:   
\[
\tilde{\rm K}^e =
\frac{1}{2}
\begin{pmatrix}
-2 & 1&1\\
1 & -1 & 0\\
1 & 0 & -1
\end{pmatrix},
\quad
\tilde{\rm M}^e =
\frac{1}{24}
\begin{pmatrix}
2 & 1&1\\
1 & 2 & 1\\
1 & 1 & 2
\end{pmatrix}.
\]
Let us assemble matrices for the square \RED{(denoted $s_1$)} composed of two triangles as shown in Figure~\ref{fig:triangle}, center. We have: 
\[
{\rm K}^{s_1} =
\frac{1}{2}
\begin{pmatrix}
-2 & 1&1&0\\
1&-2&0&1\\
1 & 0 & -2 & 1\\
0 & 1 & 1 & -2
\end{pmatrix},
\quad
{\rm M}^{s_1} =
\frac{1}{24}
\begin{pmatrix}
2 & 1&1&0\\
1 & 4 &2&1\\
1 & 2 & 4&1\\
0&1&1&2
\end{pmatrix}.
\]
For the square \RED{($s_2$)} shown in Figure~\ref{fig:triangle}, right we have: 
\[
{\rm K}^{s_2} =
\frac{1}{2}
\begin{pmatrix}
-2 & 1&1&0\\
1&-2&0&1\\
1 & 0 & -2 & 1\\
0 & 1 & 1 & -2
\end{pmatrix},
\quad
{\rm M}^{s_2} =
\frac{1}{24}
\begin{pmatrix}
4 & 1&1&2\\
1 & 2 &0&1\\
1 & 0 & 2&1\\
2&1&1&4
\end{pmatrix}.
\]
To \RED{suppress} the influence of the triangular tessellation on the equations, we consider an average of the two: 
\[
{\rm K}^{s} =
\frac{1}{2}
\begin{pmatrix}
-2 & 1&1&0\\
1&-2&0&1\\
1 & 0 & -2 & 1\\
0 & 1 & 1 & -2
\end{pmatrix},
\quad
{\rm M}^{s} =
\frac{1}{24}
\begin{pmatrix}
3 & 1&1&1\\
1 & 3 &1&1\\
1 & 1 & 3&1\\
1&1&1&3
\end{pmatrix}.
\]
Finally, replace the consistent element mass matrix ${\rm M}^s$ \RED{by its lumped counterpart ${\rm \hat M}^s$} \RED{obtained by adding its entries} along each row and putting the result on the main diagonal:
\[
{ \rm \hat M}^{s} =
\frac{1}{4}
\begin{pmatrix}
1 & 0&0&0\\
0 & 1 &0&0\\
0 & 0 & 1&0\\
0&0&0&1
\end{pmatrix}.
\]
The lumped matrix approach is often used in the finite element context to speed up the computations and is known not to lead to significant numerical errors \cite{Muftu2022-ia}. 

Assembling the equations using ${\rm K}^s$ and ${\rm \hat M}^s$ we get (\ref{eq:disc_Helm}) in the bulk and (\ref{eq: disc_boundary_eq}) on the boundary. For example, consider a domain $\Omega$ that consists of a single square. It is governed by the equation
\[
{\rm K}^su + k^2{\rm \hat M}^s {\rm u} =0.
\]
Each node of the square is a boundary node, and the lattice Neumann boundary conditions should be satisfied.
Indeed, the equation for the first node (let the nodes be indexed as shown in Figure~\ref{fig:triangle}, right) is
\[
\frac{1}{2}(u_3 + u_2) + \frac{1}{4}(\tilde{k}^2 - 4)u_1 = 0, 
\]
which is equivalent to (\ref{eq: disc_boundary_eq}) with (\ref{eq:disc_derivative},\ref{eq:lat_bound_Neum_b}). All the other cases can be considered in the same manner.
\end{document}